\def\deg{\mbox{$^{\circ}$}}

\newcommand\Eq[1]{Eq.$\:$[\,\ref{#1}\,]}
\newcommand\Fig[1]{Fig.$\:$\ref{#1}}

\newcommand\proton{$^1$H}
\newcommand\Cthteen{$^{13}$C}
\newcommand\musec{$\mu$s}

\newcommand\RFmax{$RF_{\rm max}$}

\documentclass[3p, twocolumn]{elsarticle}

\usepackage{graphicx}
\usepackage{colordvi}

\usepackage{amssymb}
\usepackage{rotating}
\usepackage{morefloats}

\usepackage{lineno}

\biboptions{sort&compress}

\journal{Journal of Magnetic Resonance}

\begin{document}

\begin{frontmatter}







\title{The Fantastic Four: A plug 'n' play set of \\ optimal control pulses for enhancing  nmr spectroscopy}

\author[label1,label2]{Manoj~Nimbalkar\corref{cor1}} \ead{manoj.nimbalkar@tum.de}\cortext[cor1]{Corresponding authors.}
\author[label1,label3]{Burkhard~Luy}
\author[label4]{Thomas~E.~Skinner} 
\author[label1,label5]{Jorge~L.~Neves}
\author[label4]{Naum~I.~Gershenzon}
\author[label1,label6]{Kyryl~Kobzar}
\author[label6]{Wolfgang~Bermel}
\author[label1]{Steffen~J.~Glaser\corref{cor1}} \ead{glaser@tum.de}

\address[label1]{Department of Chemistry, Technische Universit\"at M\"unchen, Lichtenbergstrasse 4, D-85747 Garching, Germany}
\address[label2]{Klinikum rechts der Isar, Technische Universit\"at M\"unchen, Ismaningerstrasse 22, D-81675 M\"unchen, Germany}
\address[label3]{Institut f\"ur Organische Chemie, Karlsruher Institut f\"ur Technologie, Fritz-Haber-Weg 6, 76131 Karlsruhe, Germany}
\address[label4]{Department of Physics, The Wright State University, 3640 Colonel John F. Glenn Highway, Dayton, OH 45435-0001, USA}
\address[label5]{Department of Fundamental Chemistry, Federal University of Pernambuco, Cidade Universit\'aria, Recife, PE 50740-560 Brazil}
\address[label6]{Bruker Biospin, Rheinstetten, Germany}

\begin{abstract}
We present highly robust, optimal control-based shaped pulses designed to replace all 90\deg\ and 180\deg\ hard pulses in a given pulse sequence for improved performance.  Special attention was devoted to ensuring that the pulses can be simply substituted in a one-to-one fashion for the original hard pulses without any additional modification of the existing sequence. The set of four pulses for each nucleus therefore consists of 90\deg\ and 180\deg\ point-to-point (PP) and universal rotation (UR) pulses of identical duration.  These 1~ms pulses provide uniform performance over resonance offsets of 20 kHz (\proton) and 35 kHz (\Cthteen) and tolerate reasonably large radio frequency (RF) inhomogeneity/miscalibration of $\pm 15\%$ (\proton) and $\pm 10\%$ (\Cthteen), making them especially suitable for NMR of small-to-medium-sized molecules (for which relaxation effects during the pulse are negligible) at an accessible and widely utilized spectrometer field strength of 600~MHz. The experimental performance of conventional hard-pulse sequences is shown to be greatly improved by incorporating the new pulses, each set referred to as the Fantastic Four (Fanta4).

\end{abstract}


\end{frontmatter}


\section{Introduction}
\label{Intro}
Hard rectangular pulses are the fundamental element of all multi-dimensio\-nal NMR pulse sequences.  They are simple to use and extremely versatile.  The transformations required for multi-dimensional spectroscopy---excitation, flip-back, inversion, and refocusing---are all possible using this simplest of pulses.  Moreover, the performance of a single hard pulse is approximately ideal over a reasonable range of resonance offsets (proportional to the RF amplitude of the pulse) and variation in RF homogeneity ($\pm 10\%$) relevant for high resolution spectroscopy.

However, there is considerable room for improving pulse sequence performance.  RF amplitude is limited in practice and cannot be increased to match the increased pulse bandwidth needed at higher field strengths. The problem of large chemical shift can be solved by dividing the spectral region and performing multiple experiments. But this is time consuming, and unstable samples can create additional problems. Even at lower field strengths, relatively small errors produced by a single pulse can accumulate significantly in multipulse sequences.  

\begin{figure*}[tb!hp]
\centering
	\includegraphics[width=1\columnwidth]{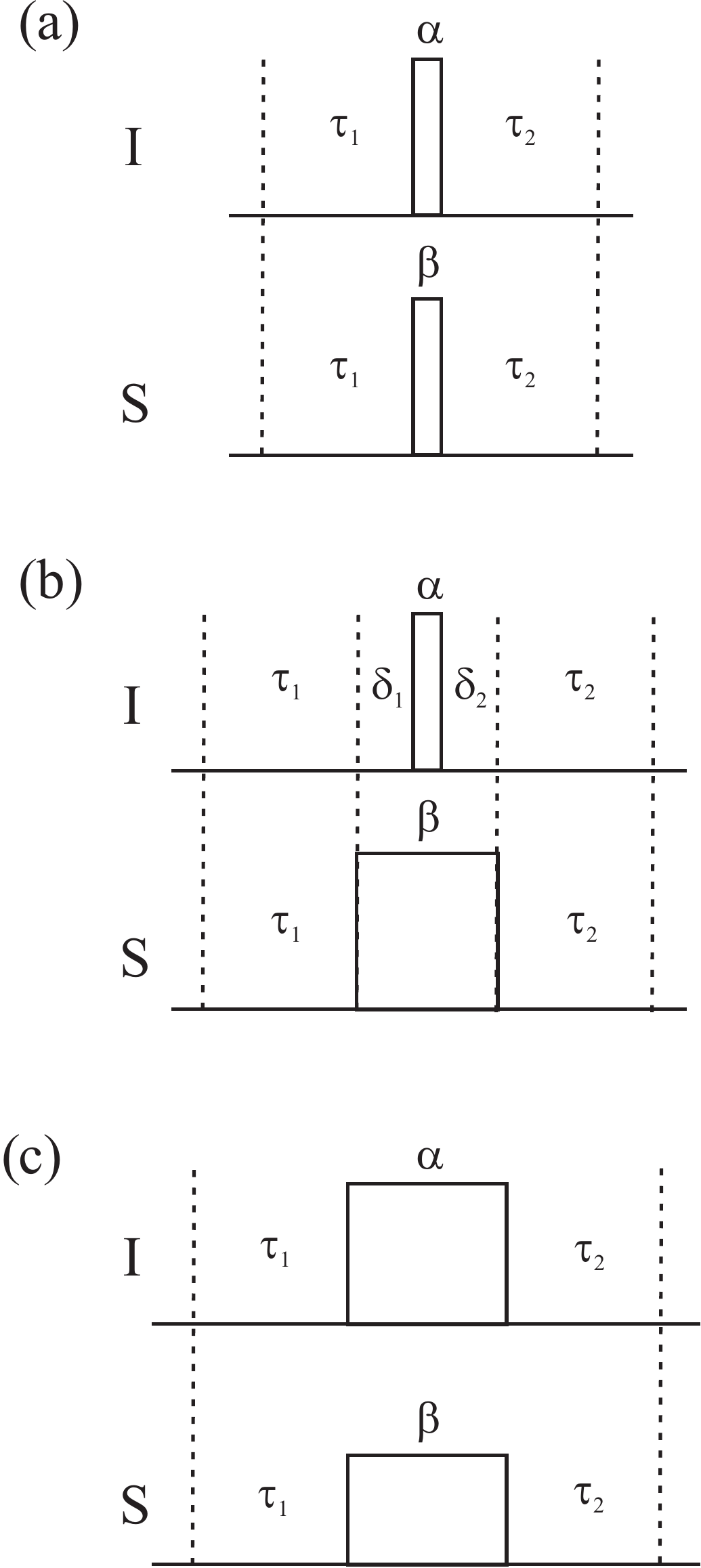}
	\caption[This schematic demonstrates a simple block of hard pulses.] {This schematic shows a pair of hard pulses with flip angles $\alpha$ and $\beta$, and delays  $\tau_1$ and $\tau_2$ (Fig.~(a)) in an any given pulse sequence. 
	If only \Cthteen\ rectangular pulses are replaced by shaped pulses (or if shaped pulses of different durations are used for \Cthteen\ and $^1$H), 
	the effect of chemical shift and coupling evolution during 
	the extra delays $\delta_1$ and $\delta_2$ have to be taken into account (Fig.~(b)).
However, if rectangular pulses are replaced by Fanta4 pulses of identical duration (Fig.~(c)), no additional evolution periods have to be taken into account.}
	\label{fig:pulse_fitting}
	\end{figure*}

The use of shaped pulses that address particular limitations of hard pulses can improve performance \cite{FKL80, M82, ME83, TCSP85, M86, SP87, BRB89, BB93, AV93, KF94, HL95, HZG97, CSS02,Hwang97, Hwang98, BEBOP1,SKLBBKG06,GSBKNLG08}, but complex multipulse sequences are masterpieces of timing and synchronized spin-state evolution.  Replacing e.g. a given pair of rectangular pulses  (c.f. Fig.~\ref{fig:pulse_fitting}a) in a sequence with a better performing pair of shaped pulses with different durations (c.f. Fig.~\ref{fig:pulse_fitting}b) typically requires a nontrivial redesign of the pulse sequence.  Additional pulses and delays are required to reestablish the timing and refocusing that achieve the goals of the original hard-pulse sequence.
In order to avoid these complications, here we developed a set of shaped pulses with {\it identical} durations, such that the replacement of rectangular pulses by shaped pulses does not introduce additional delays (c.f. Fig.~\ref{fig:pulse_fitting}c).

The goal of the present work is to provide a fundamental set of better-performing pulses that can simply replace, in one-to-one fashion, all the hard 90\deg\ and 180\deg\ pulses in any existing NMR 
sequence.  This includes important sequences such as HSQC, HMBC, HMQC, INADEQUATE, COSY, and NOESY, which are some of the most basic pulse sequences used for finding correlations between nuclei within a given molecule.  Since hard pulses perform a universal rotation (UR) about a given fixed axis for any orientation of the initial magnetization, better-performing UR 90\deg\ and UR 180\deg\ pulses are, in principle, all that are needed.  However, UR pulses are only strictly necessary for refocusing or combined excitation/flip-back. Point-to-point (PP) pulses, which transform only one specific initial state to a desired final state, are sufficient for excitation, flip-back, or inversion, and these can be designed more easily and with better performance for their specific task than UR pulses \cite{URconstruction,ToroidPulses,BURBOP}.  

We used the optimal control-based GRAPE algorithm \cite{GRAPE} to design 90\deg\ and 180\deg\ pulses for UR and PP transformations applicable to \proton\ and \Cthteen\ spectroscopy at 600~MHz, which is currently a generally accessible and widely utilized spectrometer field strength.  Pulse bandwidths are therefore 20~kHz for \proton\ (33.33~ppm) and 35~kHz for \Cthteen\ (233.33~ppm).  In order to universally implement these shaped pulses on any available probe-head, the maximum RF amplitude for \Cthteen\ spins was set to $RF_\textrm{max} = 10$~kHz, robust to $\pm$10$\%$ RF inhomogeneity/miscalibration. For \proton\ nuclei, $RF_\textrm{max} = 18$~kHz was allowed, with tolerance to $\pm$15$\%$ variations.

A pulse length $T_p = 1$~ms is the minimum length necessary to achieve suitably uniform performance for the 180\deg\ UR transformation over resonance offsets of 35~kHz with $RF_\textrm{max}$ limited to 10~kHz.  Although the same performance can be obtained using shorter pulse lengths for the other pulse criteria, all pulses were designed using the same 1~ms length.  Each hard pulse in a sequence can then be replaced by the corresponding UR or PP pulse without any further modification of the sequence.
This pulse length is most suitable for spectroscopy of small-to medium-sized molecules, where relaxation effects are neglibile during the pulse. The final product is two sets of four pulses, each set whimsically referred to as the Fantastic Four (Fanta4) due to the significant improvement they provide in pulse sequence performance.

\section{Optimization}
\label{sec:optimization}

The GRAPE algorithm for pulse optimization is discussed in detail in \cite{GRAPE}.  Further details of its application can be found in the cited references on optimal control \cite{BEBOP1,BEBOP2,BEBOP3,Limits, Pattern, MagicPulse, RC-BEBOP, ICEBERG, Limits2, OP_Algorithm}, and in particular the references on UR pulses \cite{URconstruction,  ToroidPulses, BURBOP}.  A quality factor, $\Phi$, for pulse performance is defined which, in turn, enables an efficiently calculated gradient for iterative improvement of pulse performance.  Most generally, the quality factor is a quantitative comparison between the state of the system and some desired target state.  The gradient therefore also depends on the system state which, for a single spin, is given by the magnetization $\mathbf{M}$.  Modifications in the basic algorithm that are required for two-spin systems requires some elaboration.

\subsection{Two-spin systems}
Evolution due to heteronuclear $J$-coupling during a pulse must be considered during the optimization of the Fanta4 pulses.  At the simplest level, a refocused pulse with no chemical shift evolution during the pulse produces no heteronuclear $J$-coupling evolution when applied to a single spin \cite{ICEBERG}.
However, this is not the case when pulses are applied simultaneously to \proton\ and \Cthteen, where Hartmann-Hahn transfer must be considered. In principle, this is readily addressed by optimizing shaped pulses for a coupled two spin-1/2 system simultaneously.  

However, optimization of shaped pulses for coupled two-spin systems over a range of resonance offsets and tolerance to RF inhomogeneity is computationally very expensive. For example, a one-spin system with an offset range of 20--35~kHz digitized in 250--500~Hz increments requires on the order of 100 offsets (choosing rounded numbers for simple illustration).  Tolerance to RF inhomgeneity of $\pm$10--15\% can be included using $\sim10$ RF increments, giving an order-of-magnitude total of 1000 combinations.  The time evolution and cost $\Phi$ must be calculated for each combination given a particular trial pulse in the iterative optimization procedure.  Adding a second spin, the number of combinations becomes $10^6$, increasing the computation time by 3 orders of magnitude.  The actual time can easily be 4 orders of magnitude larger than the single-spin case when all actual additional factors of 2--3 are included.  Even parallel processing (as we have utilized for a decade) requires a nontrivial amount of processing time.  
Details of the coupled two-spin-1/2 optimizations and the resulting pulses are provided in the Appendix (see section \ref{twospin}). 

The optimization algorithm begins with a trial pulse.  In the absence of a known pulse that performs reasonably well as a candidate for the optimization, the initial pulse is typically generated as a sequence of random amplitudes and phases.  Different initial pulses most generally result in different final pulses which can vary in performance, so some experimentation is necessary using multiple trial pulses to determine the ultimate quality factor that is physically attainable for a given set of performance criteria. 

\subsection{Single-spin proxy}
The physical performance limits of pulses applied to single-spin systems have been studied in detail for broadband excitation and inversion, both for fixed  $RF_\textrm{max}$ \cite{Limits} and fixed RF power \cite{Limits2}.  A complete and thorough characterization of UR pulse performance limits is also available \cite{KyrylThesis, arxivSGNBLG}. 

We used these single-spin results as a guide for what might be possible for two-spin systems.  As an alternative to multiple, lengthy two-spin optimizations, we adopted an intuitive approach based on the likelihood that pulses separately and independently optimized for \proton\ and \Cthteen\ (with different rf amplitudes) using a single-spin protocol would not satisfy the frequency (or phase modulation) matching conditions for sufficient Hartmann-Hahn transfer.

We therefore also approximated the coupled two-spin-1/2 problem as two independent, non-interacting single spin-1/2 problems, each of which demands less computation by several orders of magnitude compared to the two-spin case. We then simulated simultaneous application of the resulting pulses to \proton\ and \Cthteen\ using the full two-spin-1/2 system with a $J$ coupling of 200~Hz. 

For example, a PP 90\deg\ shaped pulse along the $x$-axis should rotate initial $z$-magnetization to the $-y$-axis.  In the case of a coupled two-spin system, a combination of a PP 90\deg\ pulse on \proton\ ($I$-spin) and \Cthteen\ ($S$-spin) should rotate initial $I_z$ and $S_z$ to $I_y$ and $S_y$ with maximum fidelity and a minimum of unwanted terms such as $2I_yS_z$ or $2I_zS_y$.  The best shaped pulses which fell within the threshold value of fidelity were included in the Fanta4 pulse set. Further details of Fanta4 pulse selection using this protocol are provided in the Appendix (section \ref{fanta4selection}).

We also note that if only one pulse is applied to either of the spins, transverse magnetization of the other spin can evolve under the chemical shift Hamiltonian. We therefore require a ``Do-Nothing'' pulse which rotates magnetization by an arbitrary multiple of 360\deg, which can be realized by two identical PP inversion pulses with a duration of 500~\musec\ each.

\section{Experiments}
\label{sec:experiments}

\begin{figure*}[tb!hp]
\centering
	\includegraphics[width=1.5\columnwidth]{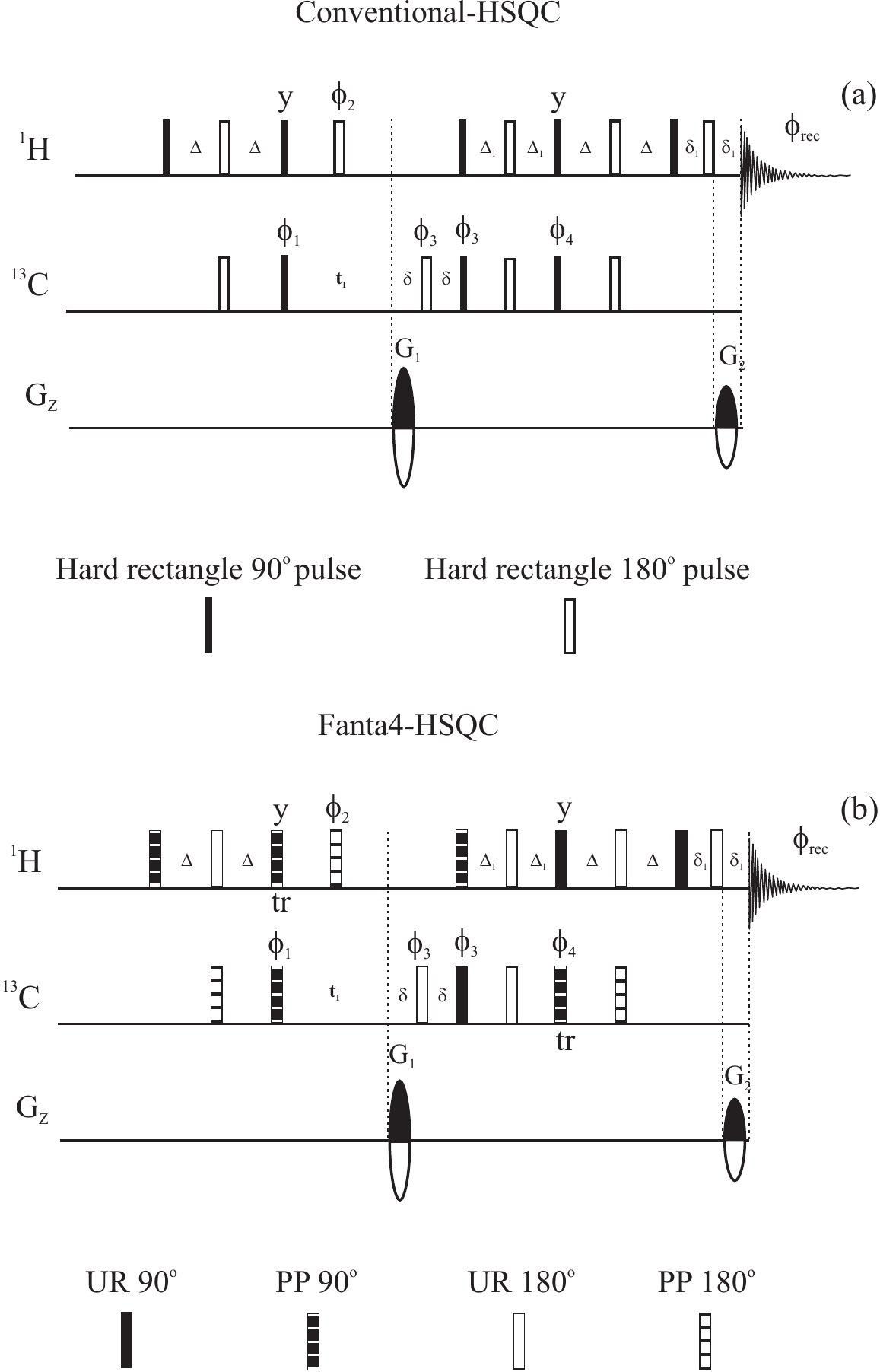}
	\caption[Creation of Fanta4-HSQC from conventional-HSQC.]{Proton excited and detected HSQC \cite{PIIICWR91, SSSSSGSG94, KKS91} experiment using conventional 90\deg\ and 180\deg\ hard pulses (conventional-HSQC, Fig.~(a)) and Fanta4 pulses (Fanta4-HSQC, Fig.~(b)). Phases are x with the exception of $\Phi_1$ = x -x, $\Phi_2$ = x x -x -x, $\Phi_3$ = x x -x -x, $\Phi_4$ = y y -y -y, $\Phi_{rec}$ = x -x -x x. Delays are $\Delta$ = 1/(4J) for CH groups and  $\Delta_1$ = 1/(8J) for all multiplicities. $\delta $ and $\delta_1$ are delays for gradients (G$_1$ and G$_2$) including recovery time. `tr' is a time reverse shaped pulse. Both pulse sequences are practically of identical lengths but the offset-compensated and RF robust Fanta4-HSQC provides higher sensitivity.}
	\label{fig:pulse_hsqc}
	\end{figure*}

\begin{figure*}[tb!hp]
\centering
	\includegraphics[width=1.6\columnwidth]{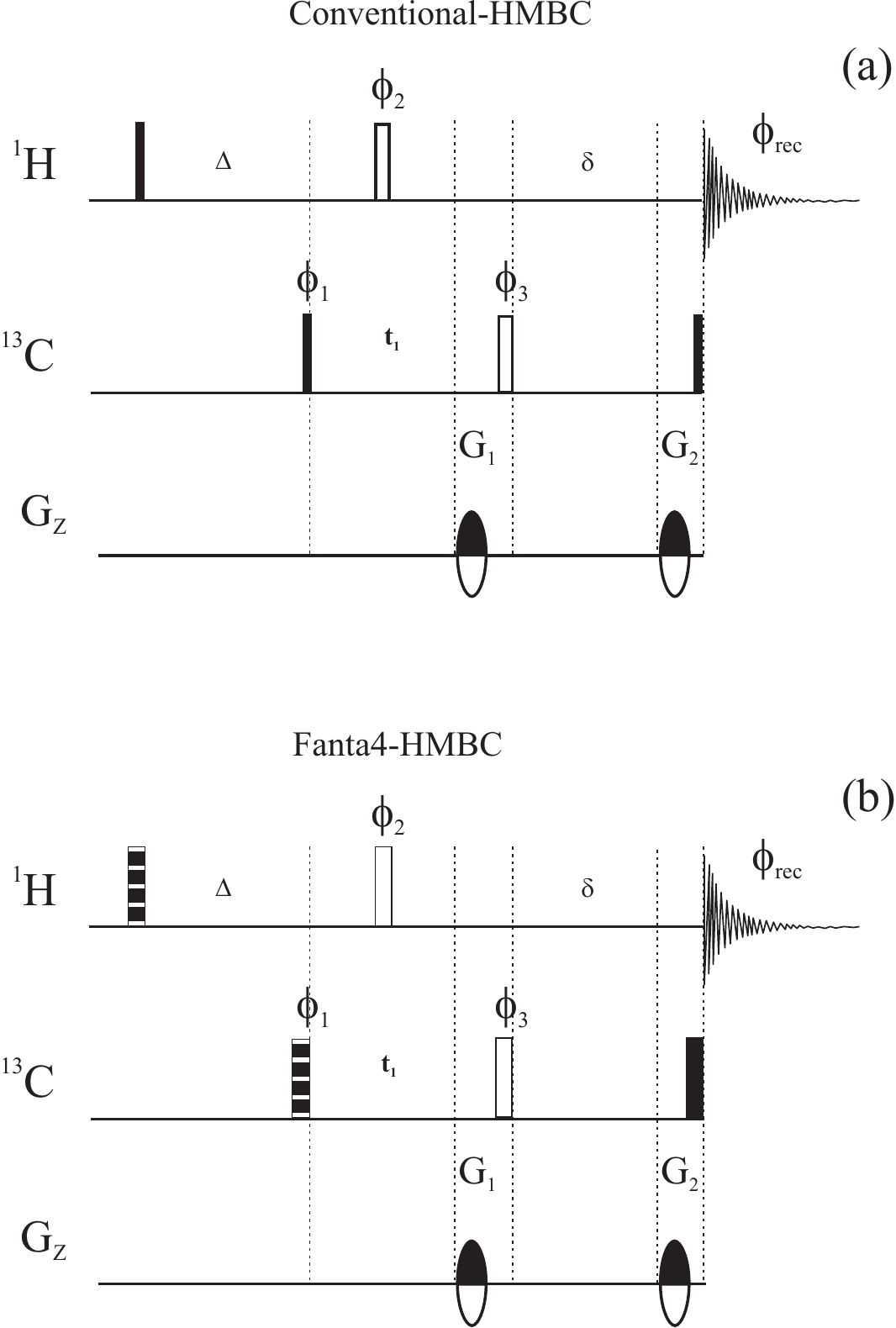}
	\caption[Creation of Fanta4-HMBC from conventional-HMBC.]{Proton excited and detected HMBC experiment \cite{CGBB01} using conventional 90\deg\ and 180\deg\ hard pulse (conventional-HMBC (Fig.~(a))) and Fanat4 pulses (Fanta4-HMBC (Fig.~(b))). Phases are x with the exception of $\Phi_1$ = x -x, $\Phi_2$ = x x -x -x, $\Phi_3$ = x x x x -x -x -x -x, $\Phi_{rec}$ = x -x x -x -x x -x x. Delays are $\Delta$ = 1/2J for evolution of long range couplings. $\delta$ is delay for gradients (G$_1$ and G$_2$) including recovery time. Both pulse sequences are practically of identical lengths but the offset-compensated and RF robust Fanta4-HMBC provides higher sensitivity.}
	\label{fig:pulse_hmbc}
	\end{figure*}

We chose the HSQC (Fig.~\ref{fig:pulse_hsqc}) and HMBC (Fig.~\ref{fig:pulse_hmbc}) experiments to demonstrate the implementation and performance advantages of Fanta4 pulses. HSQC is widely used for recording one-bond correlation spectra between two heteronuclei.  HMBC is mostly used for correlating heteronuclei connected by multiple bonds, mostly 2-4 bonds. Hard pulses were replaced by corresponding Fanta4 pulses in these pulse sequences. 
For the most efficient replacement, the task of each hard pulse in  Figs. \ref{fig:pulse_hsqc}a and ~\ref{fig:pulse_hmbc}a, is analyzed and classified according to the following four pulse types: UR 90\deg, PP 90\deg, UR 180\deg and PP 180\deg (represented in Figs. 
\ref{fig:pulse_hsqc}b and ~\ref{fig:pulse_hmbc}b
by solid black bars, black bars with white horizontal stripes, white bars and white bars with black horizontal stripes, respectively). For example, the first $^1$H 90\deg $x$-pulse acts on initial z magnetization and hence can be replaced by a PP 90$^\circ_y$  pulse that transfers $I_z$ to $-I_y$.
In the HSQC experiment, the following  $^1$H-180\deg pulse is a refocussing element that needs to be implemented as a UR 180\deg pulse, whereas the simultaneously applied $^{13}$C-180\deg pulse only needs to invert $^{13}$C spins and can be replaced by a PP 180\deg pulse.
Note that in the back transfer with sensitivity enhancement, some of the 90\deg pulses (represented by solid bars in Fig. \ref{fig:pulse_hsqc} b) also need to be implemented as UR 90\deg pulses as they simultaneously need to act on two orthogonal spin operators.

All experiments were performed at 298\deg~K on a Bruker 600 MHz AVANCE III spectromter equipped with SGU units for RF control and linearized amplifiers, utilizing a triple-resonance TXI probehead and gradients along the z-axis. 
Further details of individual pulse shapes and performance as a function of resonance offset are provided in the Appendix (see section \ref{fanta4excitaion}).

\section{Results and Discussion}
\label{sec:results}

\subsection{HSQC: Testing with Sodium Formate}
\label{subsec:testing with formate}

\begin{figure*}[tb!hp]
\centering
	\includegraphics[width=1.2\columnwidth]{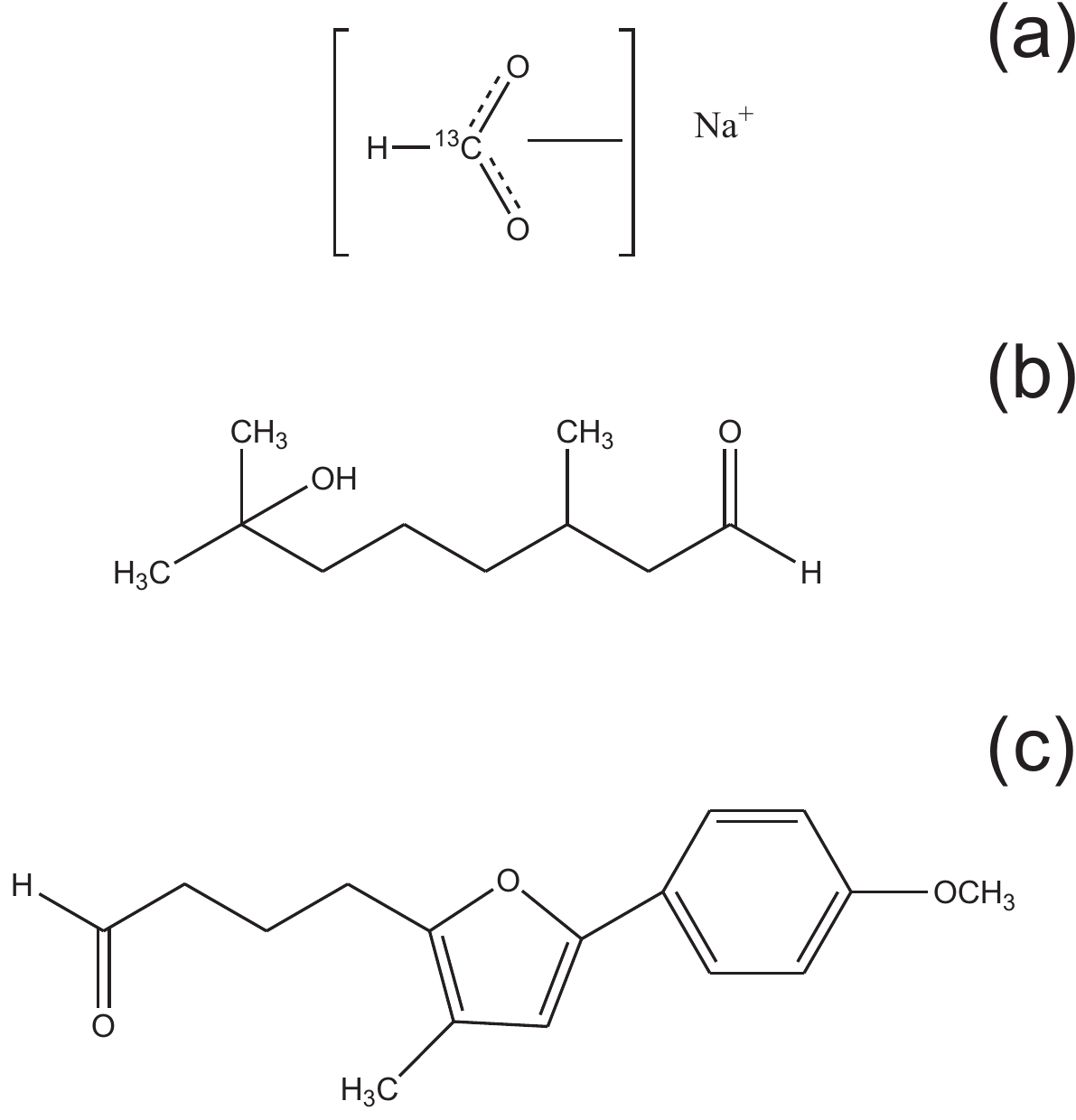}
	\caption [The structure of molecules used for comparison of  conventional pulses sequences and Fanta4 pulse sequences.]{(a) Sodium formate, (b) Hydroxycitronellal, and (c) 4-(5-(4-methoxyphenyl)-3-methylfuran-2-yl)butanal from \cite{UPHK11} are used for comparison of  conventional HSQC and HMBC pulse sequences, and Fanta4 -HSQC and -HMBC pulses sequences.}
	\label{fig:molecules}
	\end{figure*}

${^{13}}\mathrm{C}$-labeled Sodium formate (Fig.~\ref{fig:molecules}a), forming a simple two-spin system with one proton and one carbon spin ($J = 197$~Hz) dissolved in D$_{2}$O was used to compare the performance of conventional, and adiabatic-HSQC \cite{PIIICWR91, KKS91, SSSSGSG94} (where \Cthteen\  180\deg\ hard pulses were replaced by  adiabatic inversion and composite refocusing pulses constructed using Chirp60 \cite{Hwang98,Hwang97} from Bruker library \cite{Bermel03}) to the Fanta4-HSQC sequence at different chemical shifts and RF miscalibrations. 
To reduce the effects of RF field inhomogeneity, a small volume of approximately 40~$\mu$l of sample solution was placed in a 5~mm Shigemi limited volume tube.
A proton-excited and detected HSQC (coupled during acquisition) gives a proton doublet with respect to carbon.  

\begin{figure*}[tb!hp]
\centering
	\includegraphics[width=2\columnwidth]{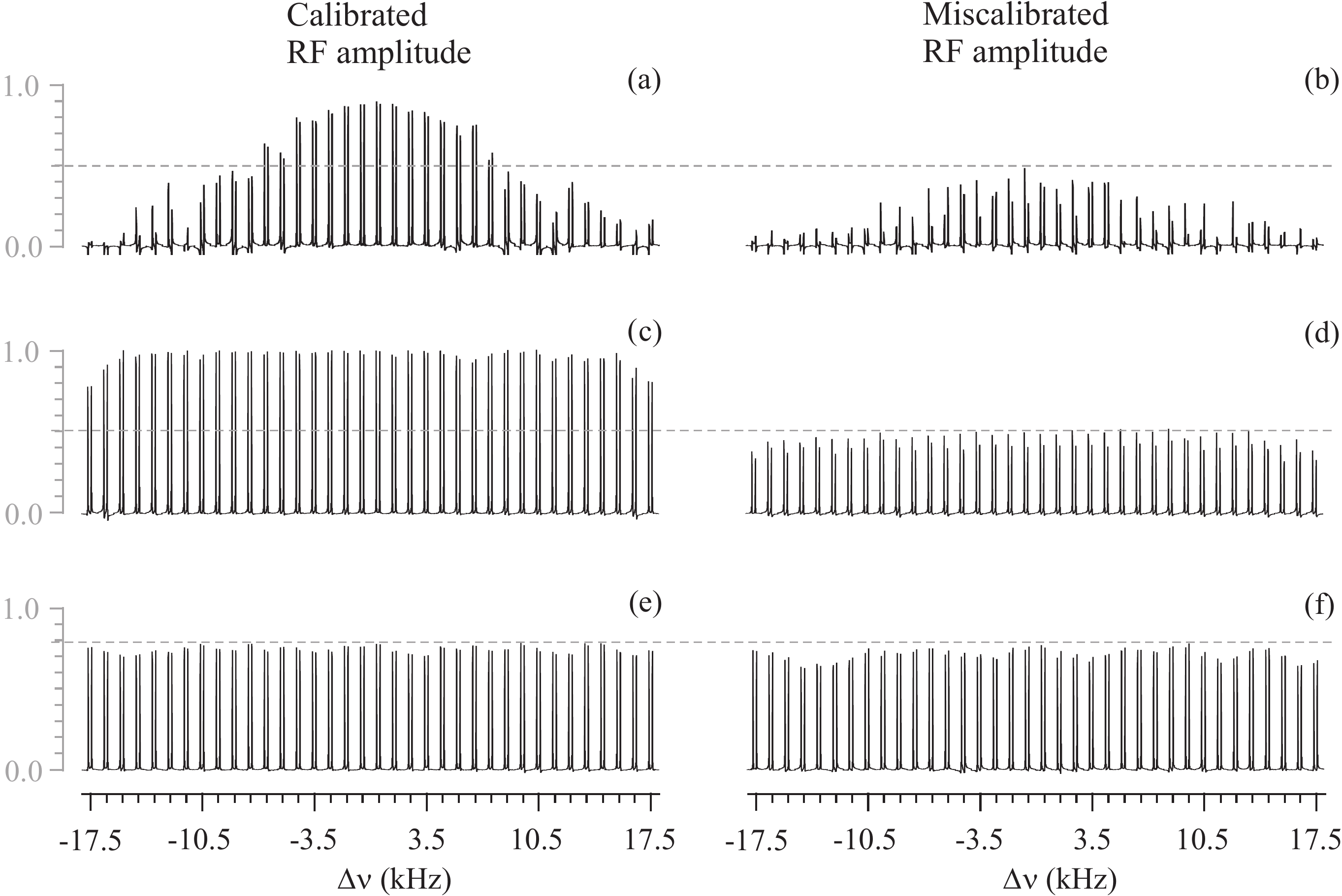}
	\caption[1D-HSQC: Chemical shift ($\Delta$v) excitation profile comparison of Fanta4 and conventional -HSQC.]{Chemical shift ($\Delta$v) excitation profile comparison of the conventional-HSQC (Fig.~(a) and (b)), the adiabatic-HSQC (Fig.~(c) and (d)), and the Fanta4-HSQC (Fig.~(e) and (f)) acquired on a \Cthteen\ labeled Sodium formate dissolved in D$_2$O. HSQC (coupled during acquisition) gives doublet of a proton with respect to a carbon. Figures~(a), (c), and (e) compare signal intensity at different \Cthteen\ resonance offsets with ideal RF amplitude, and Figures (b), (d), and (f) show the effect of RF inhomogeneity/miscalibration for ${^{1}}\mathrm{H}$ (-15$\%$) and  ${^{13}}\mathrm{C}$ (-10$\%$) pulses for different ${^{13}}\mathrm{C}$ offsets. In conventional-HSQC and adiabatic-HSQC, the ideal RF amplitude on  ${^{1}}\mathrm{H}$  hard pulses was 29.51 kHz and  ${^{13}}\mathrm{C}$ hard pulses was 19.53 kHz, while adiabatic shaped (inversion and refocusing) pulses on ${^{13}}\mathrm{C}$  were with 10 kHz maximum amplitude. In case of Fanta4-HSQC, shaped pulses on ${^{1}}\mathrm{H}$ had a maximum amplitude of 18 kHz, while on ${^{13}}\mathrm{C}$ 10 kHz.}
	\label{fig:offset_profile_hsqc}
	\end{figure*}

1D-HSQC experiments were performed incrementing the carrier frequency successively over the offset range in the \Cthteen\ dimension.  
The correctly calibrated RF amplitude is applied to \proton\ and \Cthteen\ in the comparison between conventional, adiabatic, and Fanta4 pulses shown in 
Figs.~\ref{fig:offset_profile_hsqc}a, \ref{fig:offset_profile_hsqc}c, and \ref{fig:offset_profile_hsqc}e. 
The conventional HSQC produces signal intensity within 80\% of the on-resonance signal of the calibrated pulse
over at most an offset range of $\pm 5$~kHz, far short of the desired 35~kHz bandwidth. RF pulse errors accumulate during the sequence, resulting in degraded signal intensity with considerable phase errors, in contrast to the nearly uniform performance of HSQC sequence with adiabatic 180\deg\ pulses on \Cthteen\ and the Fanta4 implentation over the 35~kHz offset range.  The small loss of signal ($\sim 10\%$) on resonance for the Fanta4 HSQC compared to the hard-pulse HSQC is more than compensated by signal gains and more uniform performance over the desired bandwidth using Fanta4. The contrast is especially striking given that the peak amplitudes of the hard pulses were higher than the Fanta4 peak amplitudes by almost a factor of two for \Cthteen\ and a factor of 1.6 for \proton.

For the case of adiabatic 180\deg\ pulses, peak amplitude was 10 kHz with durations of 2 ms for refocusing and 0.5 ms for inversion pulses.  The adiabatic HSQC was included to show the performance that is available by adjusting pulses for a specific application rather than using the Fanta4 pulses, which are designed to be applied more generically.  As noted in the Introduction, PP pulses can be designed for a specific task and performance criterion that are shorter than the corresponding UR pulses, but require additional adjustment to the pulse sequence.  In the present comparison, the adiabatic refocusing pulse applied to \Cthteen\ is twice as long as the Fanta4 UR 180\deg\ pulse, resulting in the better performance shown in \Fig{fig:offset_profile_hsqc}c.  However, miscalibrating the RF amplitude on \proton\ and \Cthteen\ by $-15\%$ and $-10\%$, respectively, in the conventional/adiabatic/Fanta4 comparison of Figs.~\ref{fig:offset_profile_hsqc}b, \ref{fig:offset_profile_hsqc}d and \ref{fig:offset_profile_hsqc}e only marginally affects the Fanta4 implementation, while the conventional HSQC is almost unusable, and adiabatic HSQC performed with $\sim 50\%$ loss in signal intensity.  One could further optimize the performance of the HSQC sequence by optimizing all the \proton\ and \Cthteen\ pulses and pulse timings, which is not the focus of the current article.

\subsection{HSQC: Testing with Hydroxycitronellal}
\label{subsec:testing with  Hydroxycitronellal}

We also implemented the conventional/adiabatic/Fanta4-HSQC sequences on the more complex molecule Hydroxycitronellal (\Fig{fig:molecules}b).
(In contrast to the experiments with ${^{13}}\mathrm{C}$-labeled Sodium formate with a Shigemi tube, a normal 5 mm NMR tube was used with a sample volume of about 600~$\mu$l.)
The molecule consists of a long chain of hydrocarbons with a hydroxyl group on one end and an aldehyde moiety at other end. Hydrocarbons resonate around 19~ppm, and a carbon bonded to oxygen (aldehyde moiety) resonates at 202~ppm, which corresponds to a total \Cthteen\ offset range of 27.3~kHz on a 600~MHz spectrometer. 

\begin{figure*}[tb!hp]
	\centering
	\includegraphics[width=2\columnwidth]{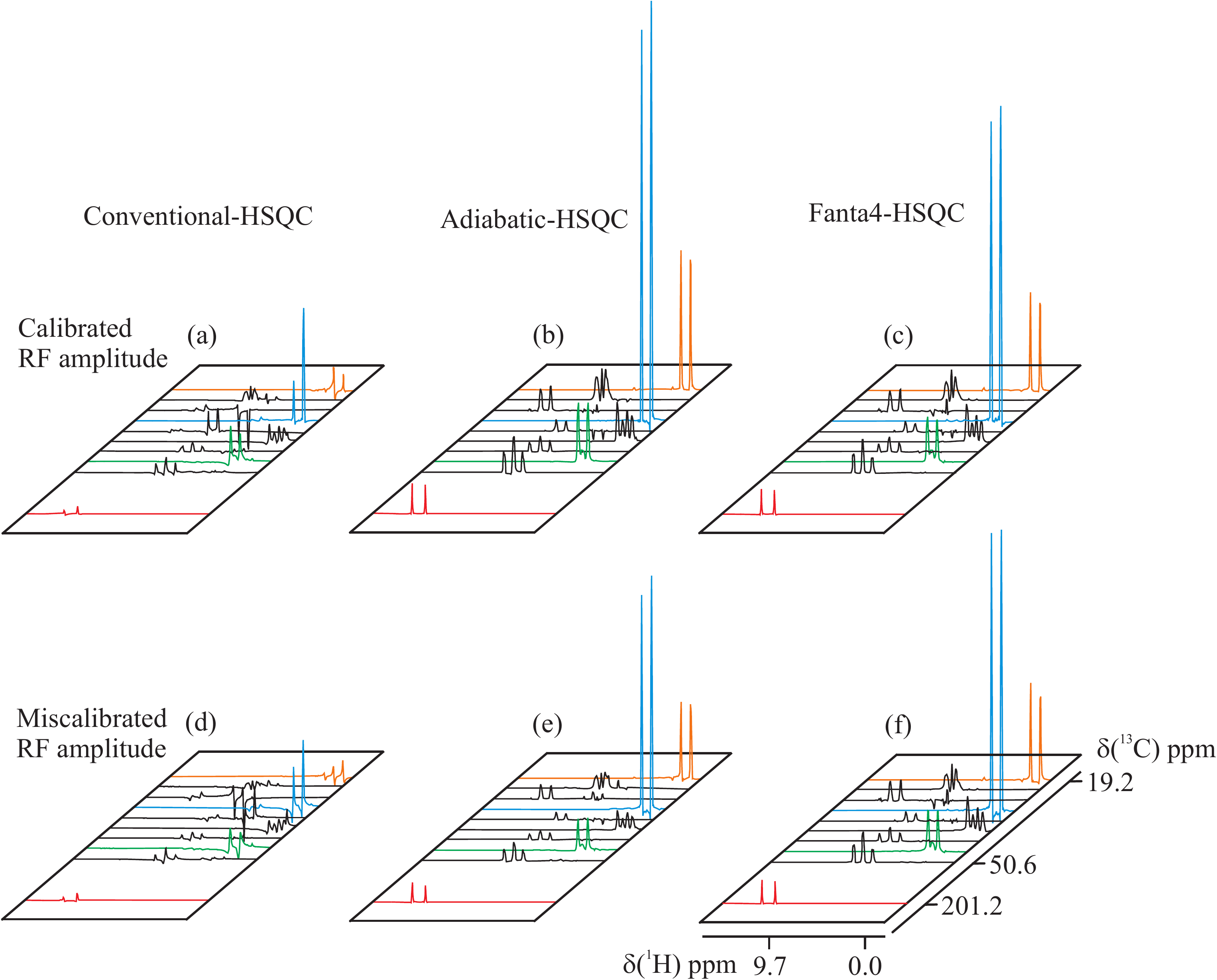}
	\caption[Fanta4-HSQC projection compared with conventional-HSQC.]{2D HSQC projection acquired on \Cthteen\ natural abundant Hydroxycitronellal, dissolved in deuterated tetrahydrofuran, using conventional (Fig.~(a) and (d)), adiabatic (Fig.~(b) and (e)), and Fanta4 (Fig.~(c) and (f)) -HSQC pulse sequences (Fig.\ref{fig:pulse_hsqc}). Figure~(a) shows the loss in signal intensity at edge of the ${^{13}}\mathrm{C}$ dimension compared to Fig.~(b), and Fig.~(c) with nominal RF amplitude. In contrast to conventional and adiabatic HSQC, only the Fanta4 HSQC is relatively unaffected by RF miscalibration of -15$\%$  on ${^{1}}\mathrm{H}$ and -10$\%$ on ${^{13}}\mathrm{C}$ pulses (compare Fig.~(d), Fig.~(e), and (f)). Amplitude on  ${^{1}}\mathrm{H}$  hard pulses was 32.26 kHz and ${^{13}}\mathrm{C}$ hard pulses was 19.53 kHz in conventional and adiabatic-HSQC. The adiabatic shaped (inversion and refocusing) pulses on ${^{13}}\mathrm{C}$  were with 10 kHz maximum amplitude. While Fanta4 pulses were with 18 kHz and 10 kHz on ${^{1}}\mathrm{H}$ and ${^{13}}\mathrm{C}$ respectively. Peaks in ${^{1}}\mathrm{H}$ and ${^{13}}\mathrm{C}$ dimension are shifted for clear distinction. $1024 \times 256$ data points were acquired with corresponding spectral widths of 230.0 ppm (${^{13}}\mathrm{C}$) and 10.0 ppm (${^{1}}\mathrm{H}$). The carrier frequency on ${^{1}}\mathrm{H}$ was 5.3 ppm and on ${^{13}}\mathrm{C}$ was 115.0 ppm. Two transients per increment gave an overall experiment time of $\sim13.5$ min for each of the three experiments.} 
	\label{2dprojection_hsqc}
	\end{figure*}

Even with
correctly calibrated RF amplitude,
the conventional HSQC experiment  is unable to excite the \Cthteen\ nuclei at very large offsets and results in poor signal-to-noise ratio in \Fig{2dprojection_hsqc}a compared to the adiabatic and  the Fanta4 HSQC shown in \Fig{2dprojection_hsqc}b and \Fig{2dprojection_hsqc}c respectively. The limitations of hard-pulse and adiabatic 180\deg\ pulse implementations of HSQC are further emphasized in the comparison between \Fig{2dprojection_hsqc}d, \Fig{2dprojection_hsqc}e and \Fig{2dprojection_hsqc}f, with the RF amplitude  miscalibrated on both nuclei as in the previous section. 

\subsection{HMBC: Testing on a complex molecule}
\label{subsec:testing with Real molecule}

A Furan with substrate 1-methoxy-4-methylbenzene (4-(5-(4-methoxyphenyl)-3-methylfuran-2-yl)butanal) (\Fig{fig:molecules}c), synthesized by a domino reaction (see table 3, entry 2 in \cite{UPHK11}), was used to test the HMBC pulse sequence.

\begin{figure*}[tb!hp]
\centering
	\includegraphics[width=2\columnwidth]{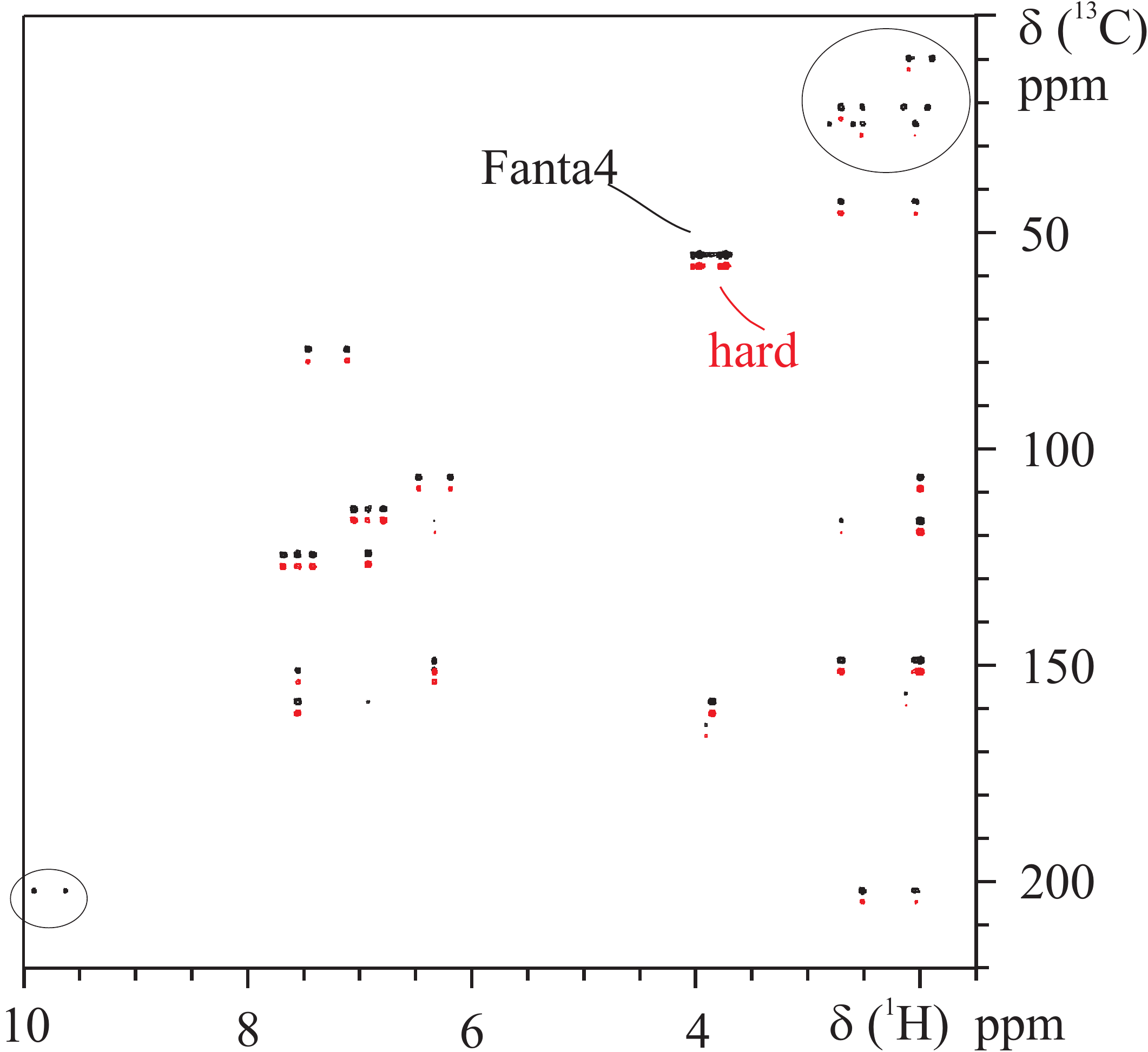}
	\caption[2D Fanta4-HMBC compared with conventional-HSQC.] {A magnitude mode 2D HMBC of a  \Cthteen\ natural abundant  4-(5-(4-methoxyphenyl)-3-methylfuran-2-yl)butanal molecule \cite{UPHK11}, dissolved in {{CDCl}}$_3$, was acquired  using conventional-HMBC (red) and Fanta4-HMBC (black) pulse sequences (Fig.~\ref{fig:pulse_hmbc}). The hard pulse version shows loss in signal intensity compared to Fanta4-HMBC(contours are slightly shifted in ${^{13}}\mathrm{C}$ dimension) near $\delta$ ${^{13}}\mathrm{C}$ = 10 to 30 ppm and at $\delta$ ${^{13}}\mathrm{C}$ = 202 ppm (encircled). $1024 \times 256$ data points were acquired with corresponding spectral widths of 230.0 ppm (${^{13}}\mathrm{C}$) and 10.0 ppm (${^{1}}\mathrm{H}$). The carrier frequency on ${^{1}}\mathrm{H}$ was 5.4 ppm and on ${^{13}}\mathrm{C}$ dimension was 115.0 ppm. Forty transients per increment gave an overall experiment time of 6 hrs 25 min for each of the two experiments. Amplitude on  ${^{1}}\mathrm{H}$  hard pulses was 30.79 kHz and ${^{13}}\mathrm{C}$ hard pulses was 19.53 kHz in conventional-HMBC,while Fanta4 amplitude were 18 kHz and 10 kHz on ${^{1}}\mathrm{H}$ and ${^{13}}\mathrm{C}$, respectively.}
	\label{fig:2d_hmbc_APII128_conv_fanta4}
	\end{figure*}

\begin{figure*}[tb!hp]
\centering
	\includegraphics[width=2\columnwidth]{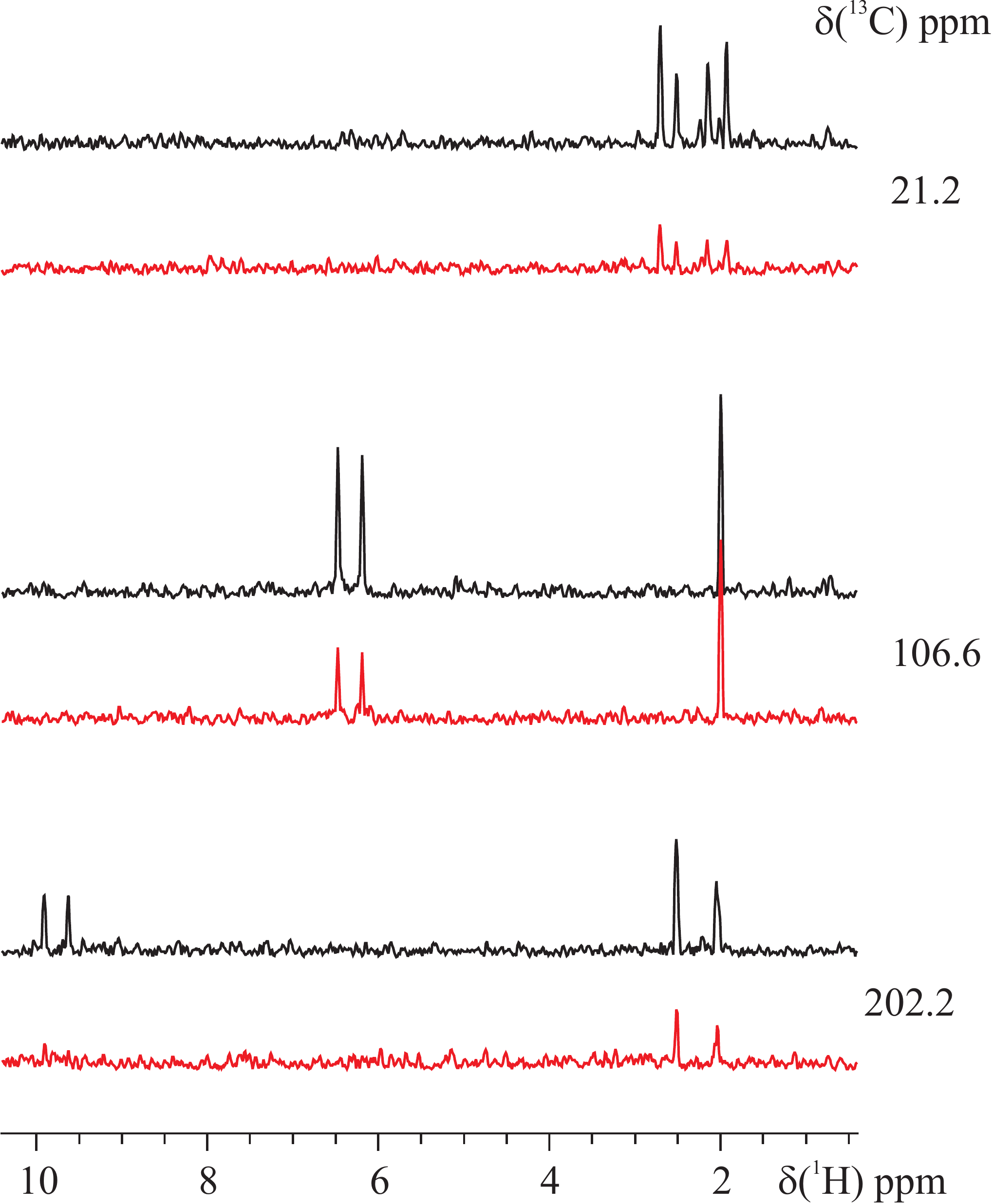}
	\caption [Traces of 2D HMBC.]{(Color online) Traces through cross signals at 21.2, 106.6 and 202.2 ppm are compared for conventional-HMBC (red, bottom) and Fanta4-HMBC (black, top) (see Fig.~\ref{fig:2d_hmbc_APII128_conv_fanta4}). The Fanta4-HMBC sequence gives improved S/N ratio compared to the conventional-HMBC sequence.}
	\label{fig:2d_hmbc_APII128_conv_fanta4_cross_section}
	\end{figure*}

 It is a relatively small molecule with a large number of long-range $J$ couplings between \proton\ and \Cthteen\ spins. The required \Cthteen\ bandwidth is 28.8~kHz at 600~MHz.
The HMBC pulse sequence with hard pulses 
(red) shows poor performance at large offsets compared to the Fanta4-HMBC (black) (Fig.~\ref{fig:2d_hmbc_APII128_conv_fanta4}). Traces from this 2D HMBC spectrum are compared in \Fig{fig:2d_hmbc_APII128_conv_fanta4_cross_section}, providing further detail of the signal-to-noise enhancement available using Fanta4, showing gains of up to a factor of three in signal-to-noise ratio.

\section{Conclusion}
\label{Conclusion}
We have designed two sets of shaped pulses, referred to as Fanta4 pulses, that can be used to replace all 90\deg\ and 180\deg\ hard pulses on \proton\ and \Cthteen, respectively, in conventional pulse sequences for improved performance.  HSQC and HMBC experiments were provided to show that each hard pulse in a sequence can be easily replaced with the corresponding Fanta4 shaped pulse without further modifying the existing pulse sequence. Compared to rectangular pulses, Fanta4 pulses provide far more robust performance with respect to large frequency offsets, RF inhomogeneity and/or RF miscalibration. The duration of the current generation of Fanta4 pulses is 1 ms, which renders them suitable for small and medium sized molecules with moderate relaxation values. 

\section{Outlook}
\label{Outlook}
Shorter pulses:
Current Fanta4 pulses are 1ms long, making them most suitable for spectroscopy of small-to medium-sized molecules, where relaxation effects are neglibile during the pulse. For applications to large molecules with considerable relaxation effects, we are in the process of designing shorter pulses.

Two-spin optimizations: 
The two-spin optimization process detailed in the Appendix~(\ref{twospin}) showed promising results. To address the previously noted problem of significantly increased computation time, we will consider a combination of a new optimization algorithm(s) and parallel programing.

\section{Appendix}

As discussed in the main text, the sets of Fanta4 pulses presented for $^1$H and $^{13}$C spins were optimized assuming single, uncoupled spins. 
In section \ref{fanta4selection}, we first explain the post-selection protocol that was used to conduct a thorough search and select the best combination of pulses (in terms of their performance for coupled spins) that were individually optimized for uncoupled spins.
In section \ref{twospin}, we discuss the far more time consuming approach of direct pulse optimization for coupled spins where spin-spin couplings are explicitly taken into account. Due to its extreme computation demands, this procedure was only used to develop a representative pulse.

\subsection{Fanta4 pulse selection}
\label{fanta4selection}

Here we explain the steps of the post-selection protocol  that we used to select  from a large set of  $^1$H and $^{13}$C PP and UR pulses that were found based on single-spin optimizations to obtain the final set of Fanta4 pulses presented in this paper. Finding the best combination of pulses based on their simulated performance in coupled two-spin systems is a nontrivial combinatorial problem which becomes tractable using the following procedure. 

\begin{sidewaysfigure*}[tb!hp]
	\centering
	\includegraphics[width=1\columnwidth]{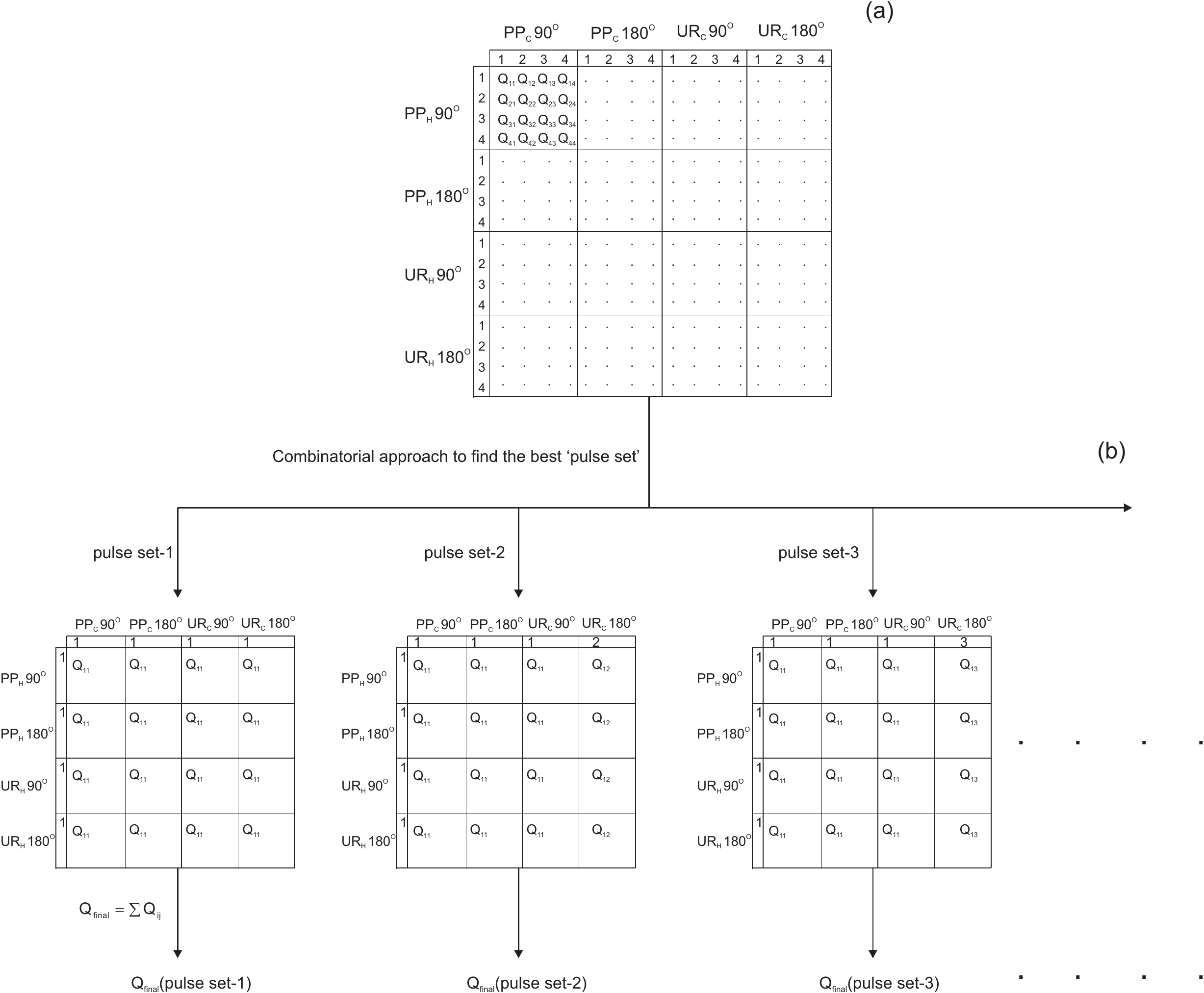}
	\caption [Figure depicts the combinatorial approach to select the best set of pulses.]{The schematic depicts the combinatorial approach to select the best set of pulses. Figure (a) lists the selected pulses with fidelity ${Q_{ij}}$, where i and i are indices for the pulse combination. Figure (b) shows the possible sets of pulses and  calculation of  the combined fidelity ${Q_{final}}$ for each pulse set.} 
	\label{fig:ps}
	\end{sidewaysfigure*}

1.
From a pool of hundreds  of optimized $^1$H and $^{13}$C~PP~90\deg, PP~180\deg, UR~90\deg, and UR~180\deg  
\ pulses (optimized starting from different random sequences)  with a duration of 1 ms, between four and 13 individual pulses 
with fidelity above 0.9999 for uncoupled spins were chosen as final candidates for each class of $^1$H and $^{13}$C PP 90\deg, PP 180\deg, UR 90\deg, and UR 180\deg \ pulses. The set of final candidate pulses is shown schematically in Fig.~\ref{fig:ps}a (assuming for simplicity four pulses in each class).

2. 
For each pair of candidate $^1$H and $^{13}$C pulses, 
the evolution of the fifteen orthogonal initial density operators $\rho^{(j)}(0)$ representing a two-spin-1/2 system (given by the
Cartesian basis operators
$\rho^{(1)}(0)=I_x$, $\rho^{(2)}(0)=I_y$, $\rho^{(3)}(0)=I_z$, 
$\rho^{(4)}(0)=S_x$, $\rho^{(5)}(0)=S_y$, $\rho^{(6)}(0)=S_z$, 
$\rho^{(7)}(0)=2 I_x S_x$, $\rho^{(8)}(0)=2 I_y S_x$, $\cdots$ $\rho^{(15)}(0)=2 I_z S_z$)
was simulated
assuming a $J$-coupling of 200~Hz and a nominal RF amplitude of 18 kHz and 10 kHz for \proton \ and \Cthteen \ pulses, respectively.
This is repeated for all combinations of  81 offsets $\nu^H$ and 141 offsets $\nu^C$ resulting from a digitization of the
 \proton \ offset range of 20 kHz and the \Cthteen \ offset range of 35 kHz  in steps of 250 Hz.

3.
For each combination of the four   \proton \ and \Cthteen \ classes, we partitioned the resulting 15 Cartesian product operators for a two-spin system in three groups: {\it desired} terms (D), {\it undesired} terms (U) and terms which can be {\it ignored} (I).
As the individual pulses have excellent performance for single spins, special attention is given here to potential adverse effects due to the $J$ coupling, such as (partial) Hartmann-Hahn transfer  in the two-spin system.

For example, consider the case of the simultaneous application of a PP 90\deg\  \proton\ (spin $I$) pulse (transferring $I_z$ to $-I_y$ for an uncoupled spin $I$) and a PP 90\deg\  \Cthteen\ (spin $S$) pulse 
(transferring $S_z$ to $-S_y$ for an uncoupled spin $S$) in the presence of $J$ coupling. If the initial density operator term is e.g.  $\rho^{(3)}(0)=I_z$, obviously the desired final operator is $-I_y$. Hence, the expectation  value of the operator $I_y$ is considered to be a {\it desired} term  (D), taking into account the correct phase according to the specification of the pulse.

The expectation values of all other terms should be as small as possible and hence are {\it undesired} (U).
However, note that a PP 90\deg\  \proton\ (spin $I$) pulse that transfers $I_z$ to $-I_y$ does not necessarily leave $I_x$ invariant - in fact the improved performance of PP 90\deg\ pulses results from the additional degree of freedom related to the fact that the final term can be anywhere in the $x$-$z$ plane. 
Hence, if the initial density operator term is $\rho^{(1)}(0)=I_x$, the expectation  values of both $I_x$ and $I_z$ are arbitrary and will not play a role in the pulse sequence, otherwise a UR pulse should have been chosen at this point in the sequence.
Therefore, for  $\rho^{(1)}(0)=I_x$, the expectation  values of the operators $I_x$  and $I_z$ can be {\it ignored} (I),
whereas all remaining operators are {\it undesirable} (U) as the initial $I_x$ operator should not leak to any of the remaining 13 product operator terms as a result of coupling evolution.

\begin{figure*}[tb!hp]
	\centering
	\includegraphics[width=1.4\columnwidth]{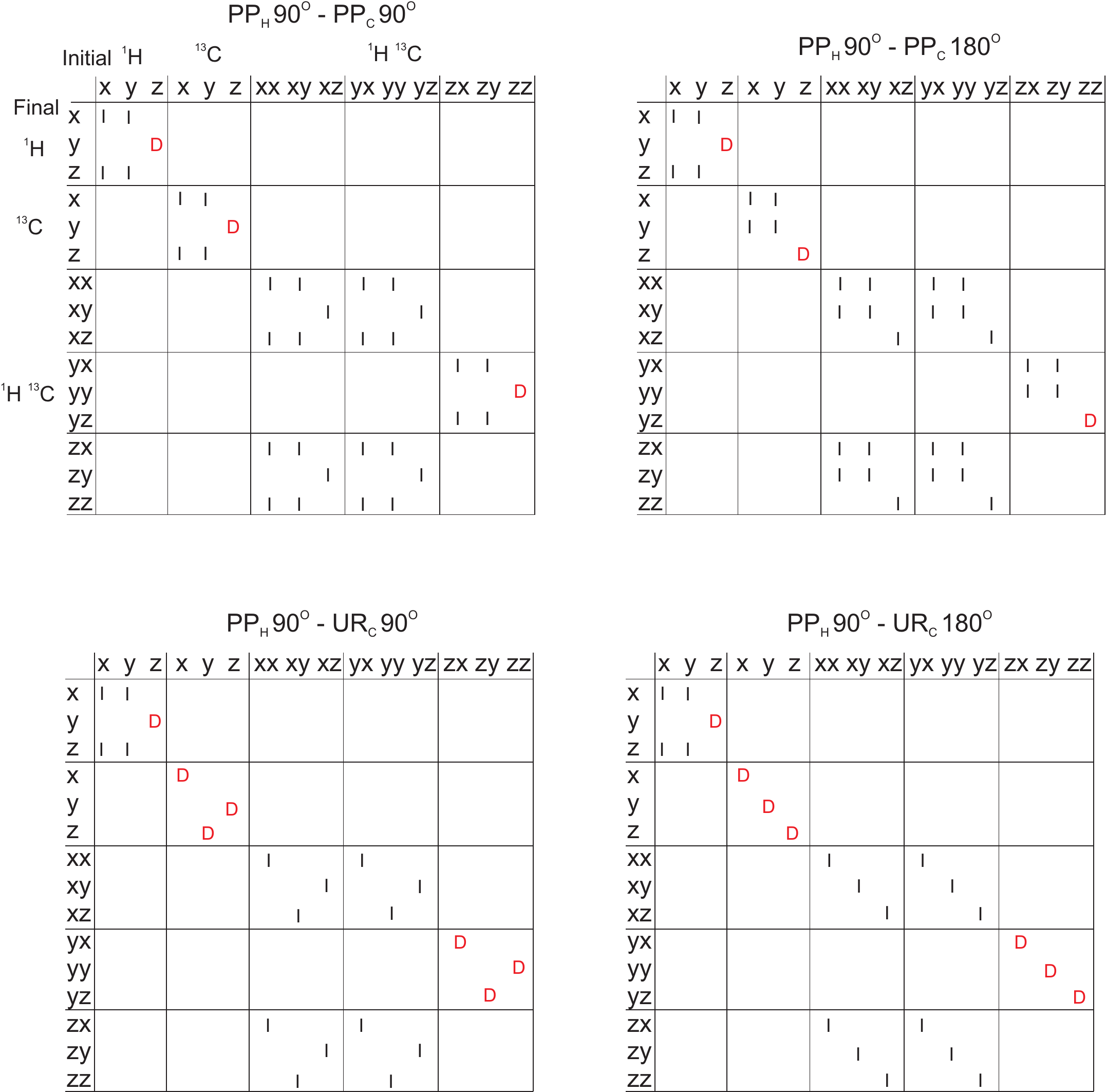}
	\caption [A product operator analysis for the combinations of $\rm{PP_H}~90\deg$ pulse with rest of \Cthteen\ pulses. ]{(Color online) This figure shows the combinations of $\rm{PP_H}~90\deg$ pulse with rest of \Cthteen\ pulses. For simulation of pair of pulses on coupled two spins-1/2 system, each component of cartesian product operator is considered as initial state and all components are detected. The desired(D, in red), undesired(empty box), and terms which can be ignored (I) are indicated in each case and described more fully in section \ref{fanta4selection}.}
	\label{fig:po_pc_l1}
	\end{figure*}

\begin{figure*}[tb!hp]
	\centering
	\includegraphics[width=1.4\columnwidth]{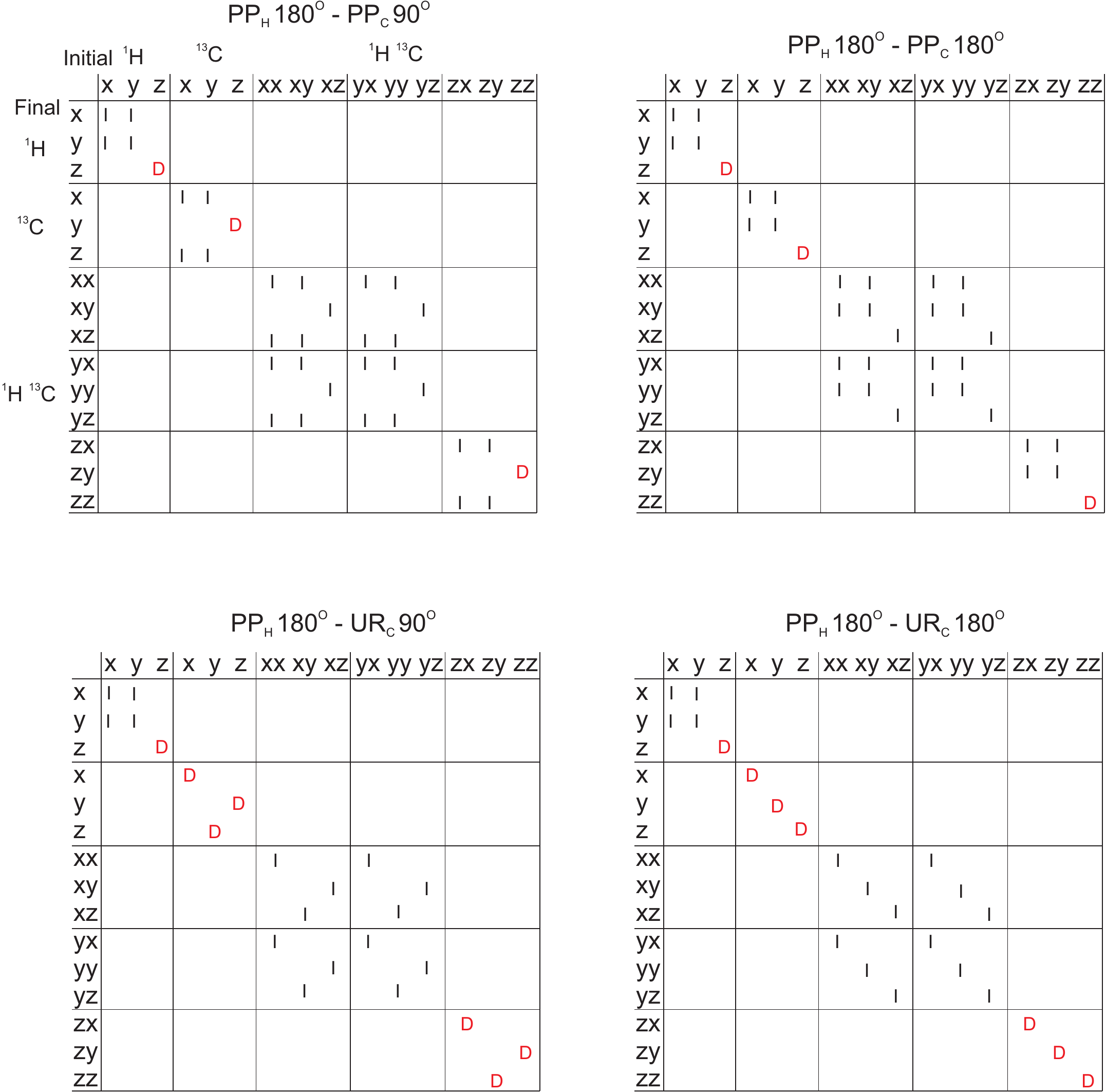}
	\caption[A product operator analysis for the combinations of $\rm{PP_H}~180\deg$ pulse with rest of \Cthteen\ pulses. ]{The combinations of $\rm{PP_H}~180\deg$ pulse with the \Cthteen\ pulses. Compare Figure ~\ref{fig:po_pc_l1}.}
	\label{fig:po_pc_l2}
	\end{figure*}

\begin{figure*}[tb!hp]
	\centering
	\includegraphics[width=1.4\columnwidth]{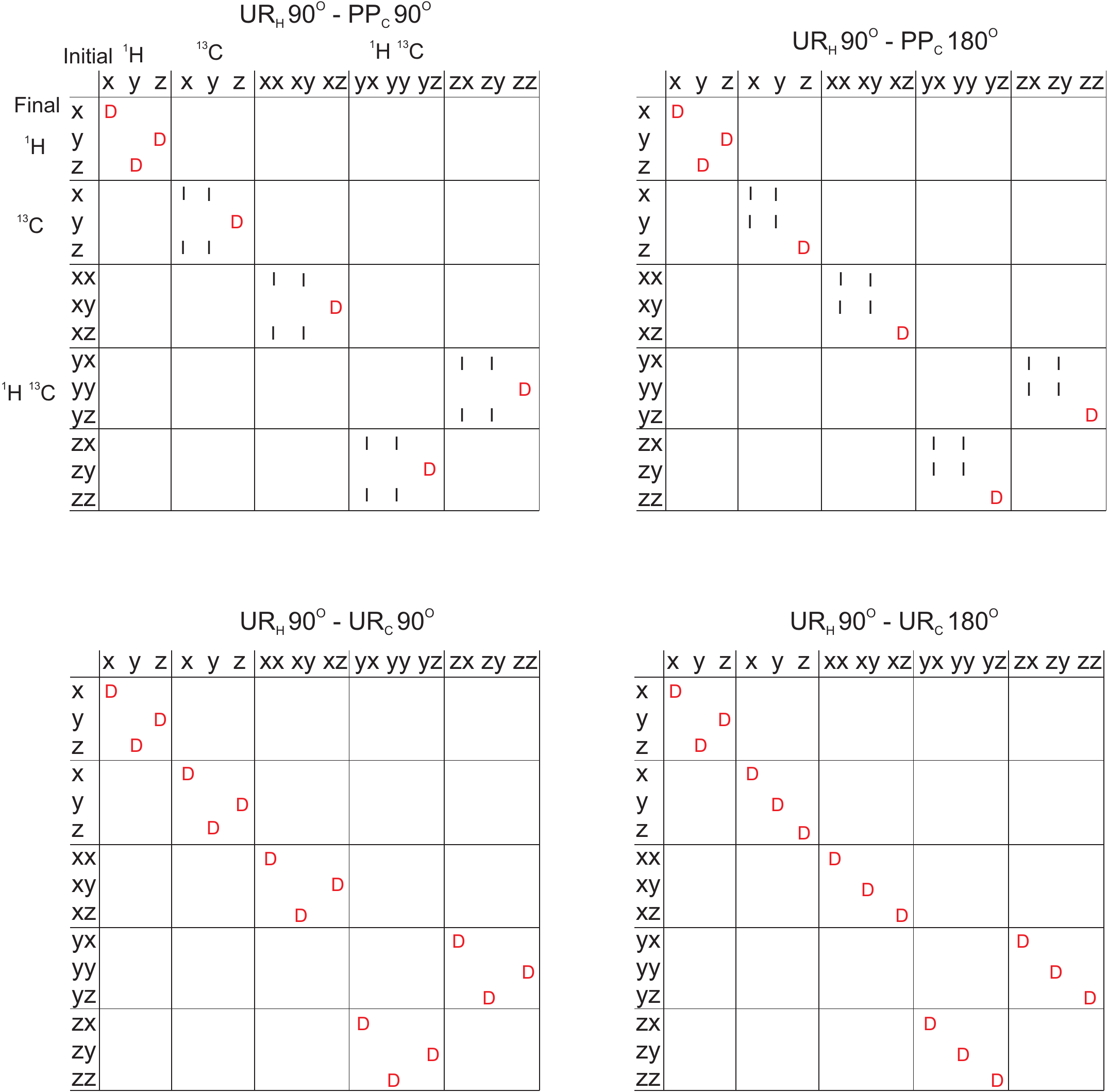}
	\caption[A product operator analysis for the combinations of $\rm{UR_H}~90\deg$ pulse with rest of \Cthteen\ pulses. ]{The combinations of $\rm{UR_H}~90\deg$ pulse with the \Cthteen\ pulses. Compare Figure~\ref{fig:po_pc_l1}.}
	\label{fig:po_pc_l3}
	\end{figure*}

\begin{figure*}[tb!hp]
	\centering
	\includegraphics[width=1.4\columnwidth]{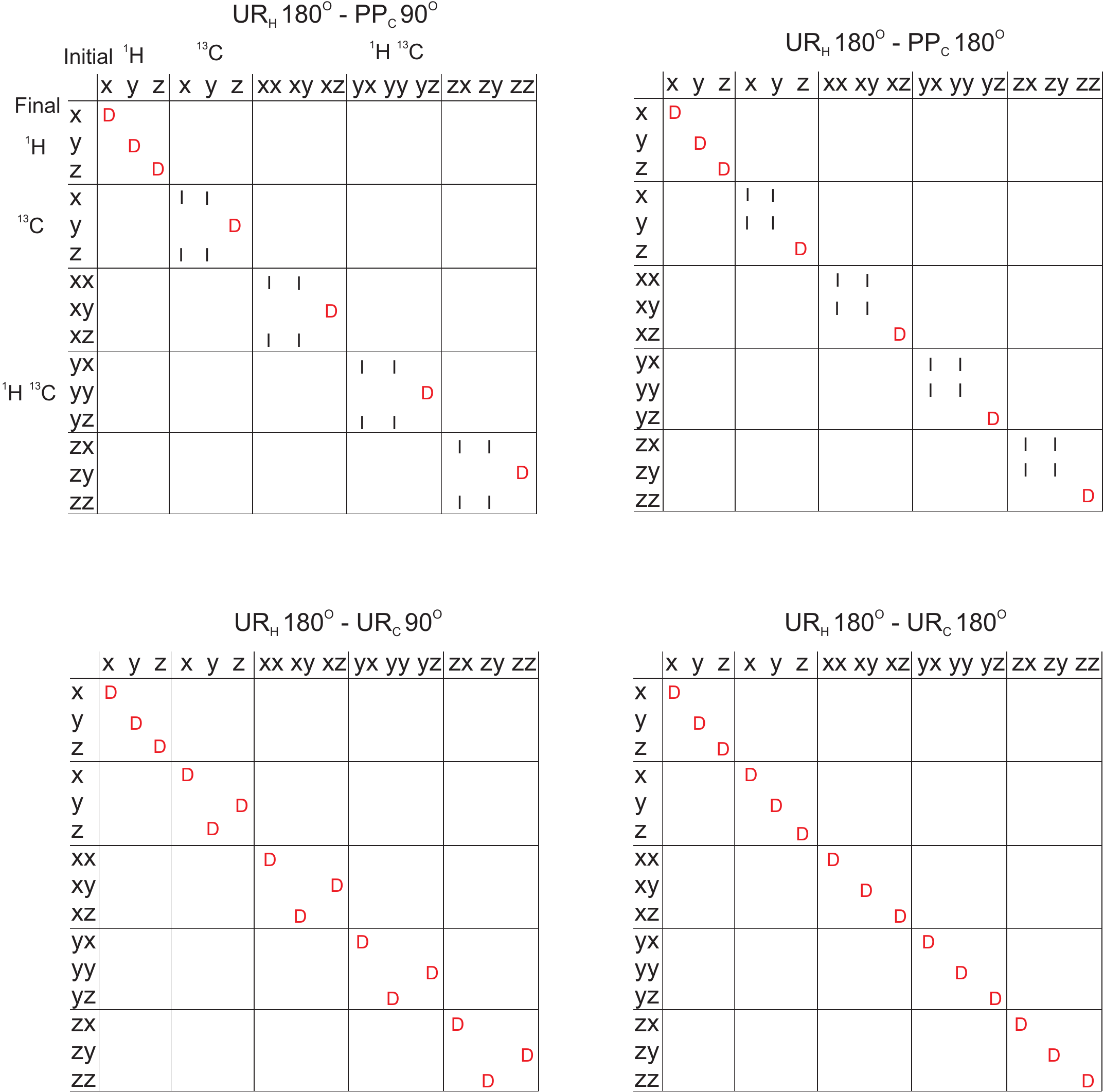}
	\caption [A product operator analysis for the combinations of $\rm{UR_H}~180\deg$ pulse with rest of \Cthteen\ pulses. ]{The combinations of $\rm{UR_H}~180\deg$ pulse with the \Cthteen\ pulses. Compare Figure~\ref{fig:po_pc_l1}.}
	\label{fig:po_pc_l4}
	\end{figure*}

For all combinations of \proton \ 
and \Cthteen \ pulse classes, the tables shown in Figs.~\ref{fig:po_pc_l1} -  \ref{fig:po_pc_l4} summarize  the desired terms (D), the undesired terms (U), and the terms of the final density operator that can be ignored (I) for all considered initial Cartesian product operators
$\rho^{(1)}(0)$,  $\cdots$ $\rho^{(15)}(0)$.

With this classification of expectation values into terms that are desired, undesired or that can be ignored,
the following quality factor is calculated for each pair of  the candidate $^1$H and $^{13}$C pulses to quantify their combined performance in the presence of the $J$ coupling:
\begin{equation}
Q_{ij}=\overline {Q_{ij}^{(D)} } - \overline {Q_{ij}^{(U)} },
\label{eq_Q}
\end{equation}
where 
${Q_{ij}^{(D)} }$ and ${Q_{ij}^{(U)} }$ are the individual quality factors    for the desired (D) and undesired (U) terms for this pulse pair

and
$\overline {Q_{ij}^{(D)} }$ and $\overline {Q_{ij}^{(U)} }$ are the averages of ${Q_{ij}^{(D)} }$ and ${Q_{ij}^{(U)} }$ 
over all combinations of offsets $\nu^H$ and $\nu^C$.
The quality factor $\overline {{Q_{ij}^{(D)} }}$ is defined as 
as the average of the absolute values of the expectation values for the desired transformations given in the corresponding table.
For example, if  the candidate pulse $i$ is
a PP 90\deg \ for $^1$H  and pulse 
$j$ is
a PP 90\deg \ for $^{13}$C, the corresponding table is found in Fig.~\ref{fig:po_pc_l1} (top left) and in this specific case
\begin{equation}
\overline {{Q_{ij}^{(D)} }} = {{(\vert \langle I_y \rangle^{(3)} \vert + \vert \langle S_y \rangle^{(6)}  \vert
+\vert \langle 2I_y S_y\rangle^{(15)} \vert} \over 3}, \ \ \
\label{eq_Qd}
\end{equation}
where $\langle A \rangle^{(p)} = {\rm Tr}\{A^\dagger  \rho^{(p)}(T)\}$ with $\rho^{(3)}(0)=I_z$ as defined above in step 2.
Similarly, the quality factor $\overline {{Q_{ij}^{(U)} }}$ is defined as the average of the absolute values of the expectation values for the undesired transformations given in the corresponding table.

Using Eq. (\ref{eq_Q}), we can calculate the quality factors $Q_{ij}$ for all pulse pairs $ij$ and fill the master table represented in Fig. 
\ref{fig:ps}a.

4. In this step,  all possible sets of Fanta4 pulses (consisting of one  PP 90\deg, one PP 180\deg, one UR 90\deg, and one UR 180\deg for both
$^1$H and $^{13}$C) are constructed
and the overall performance of each set is quantified (c.f.Fig.~\ref{fig:ps}b).
The quality factor for each set is simply given by the average of the quality factors $Q_{ij}$ of all pulse pairs in the set, where the values of 
$Q_{ij}$ have been calculated in step 3 and can be taken from the master table (Fig.~\ref{fig:ps}a).
Finally, the set with the best overall quality factor was chosen as the set of Fanta4 pulses presented in this paper.
The individual quality factors $Q_{ij}$ for the best set of Fanta4 pulses are summarized in Table 1.

The set of best pulses reduces the effect of heteronuclear J coupling during the shaped pulse.

\begin{table*}[tb!hp]
\centering
\caption[${Q}$ of is summarized for the current set of \proton\ and \Cthteen\ Fanta4 pulses.]{${Q}$ from \Eq{eq_Q} is summarized for the current set of \proton\ and \Cthteen\ Fanta4 pulses.}
\label{tab:tab1}
		\begin{tabular}{c| c c c c}
		\hline \hline
		${Q}$ & PP$_{\rm{C}}~90\deg\ $ & PP$_{\rm{C}}~180\deg\ $ & UR$_{\rm{C}}~90\deg\ $ & UR$_{\rm{C}}~180\deg\ $ \\
		\hline
		PP$_{\rm{H}}~90\deg\ $  & 0.9603 & 0.9762 & 0.9725 & 0.9679\\
		PP$_{\rm{H}}~180\deg\ $ & 0.9796 & 0.9767 & 0.9717 & 0.9701\\
		UR$_{\rm{H}}~90\deg\ $  & 0.9832 & 0.9753 & 0.9710 & 0.9634\\
		UR$_{\rm{H}}~180\deg\ $ & 0.9786 & 0.9753 & 0.9657 & 0.9665\\
		\hline \hline
		\end{tabular}
\end{table*}

\subsubsection{Fanta4 pulse shapes and excitation profiles}
\label{fanta4excitaion}

\begin{figure*}[tb!hp]
\centering
	\includegraphics[width=2\columnwidth]{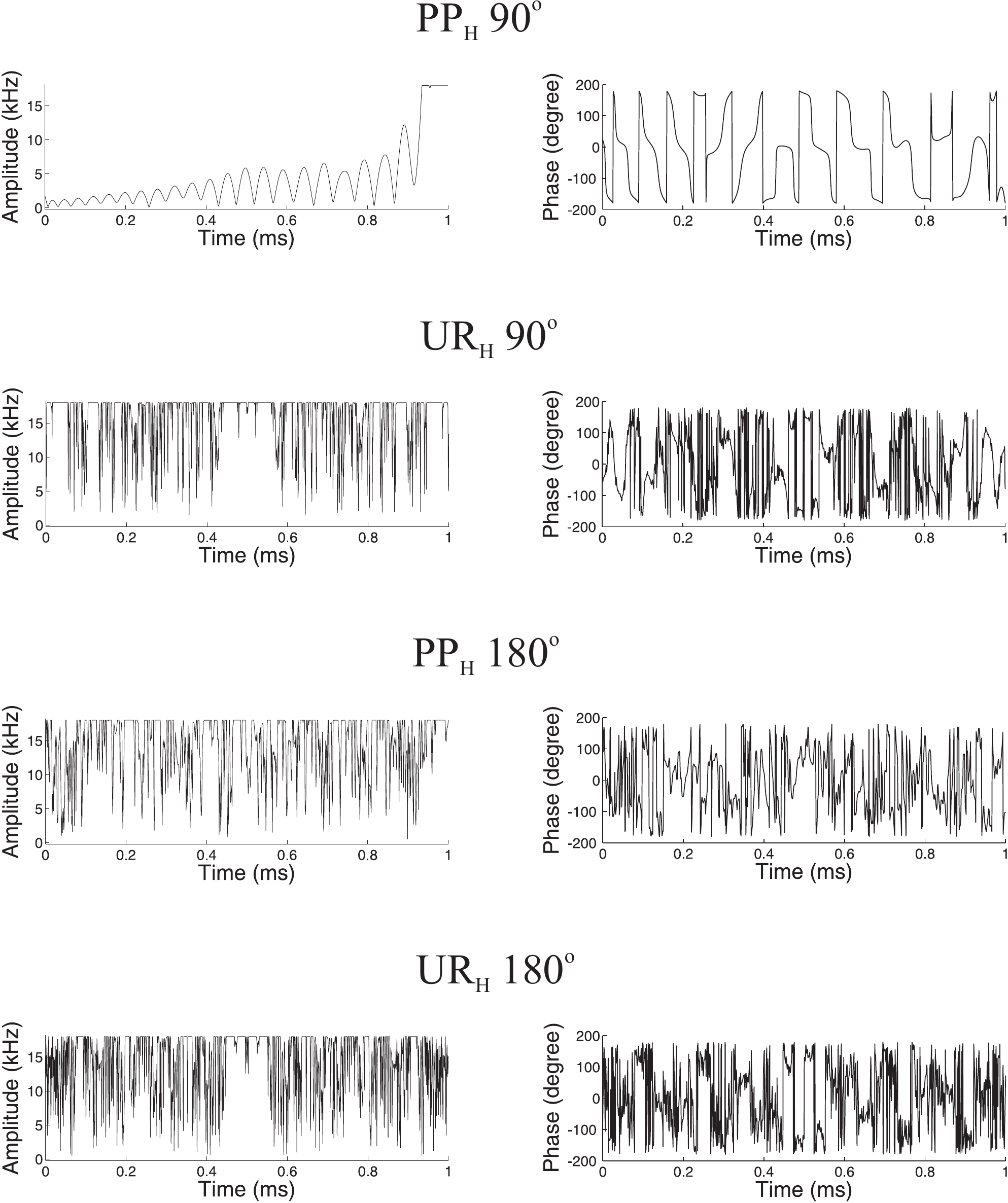}
	\caption [The amplitude and phases of \proton\ Fanta4 pulses.]{The amplitude and phase of \proton\ pulses are plotted as a function of the pulse duration. All are 1~ms long with a maximum amplitude of 18 kHz.}
	\label{fig:fanta_H_pulse}
	\end{figure*}

\begin{figure*}[tb!hp]
\centering
	\includegraphics[width=1.6\columnwidth]{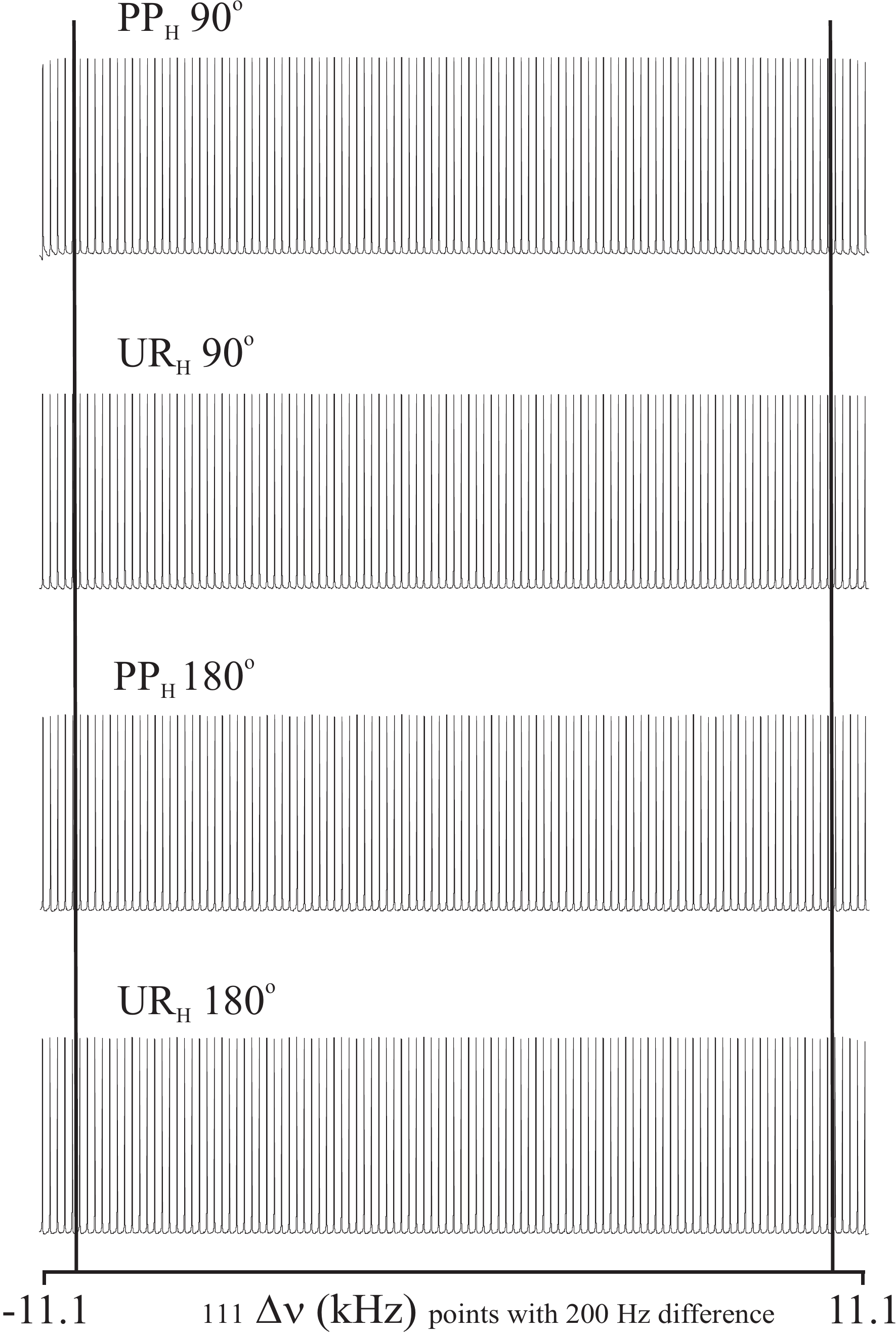}
	\caption[Experimental performance of \proton\ Fanta4 pulses as function of the offset for single spin.] {The experimental performance of \proton\ pulses is plotted as function of 
the offset $\Delta{\nu}$ at ideal RF amplitude with a \RFmax\ of 18 kHz. Experiments were 
aquired on residual proton in HDO for an offset range of 22.2 kHz in steps of 200 Hz. 
\proton\ pulses were optimized for 20 kHz of total offset range with RF miscalibration of 
$\pm5\%$, but the resulting pulses are robust for $\pm15\%$ RF miscalibration. \proton\ PP 90\deg\ and UR 
90\deg\ show the excitation profile for the $z \rightarrow -y$ and \proton\ PP 180\deg\ and UR 180\deg\ show an inversion profile from $z \rightarrow -z$.}
	\label{fig:1d_excitation_profile_fanta_H_pulse}
	\end{figure*}

\begin{figure*}[tb!hp]
	\centering
	\includegraphics[width=2\columnwidth]{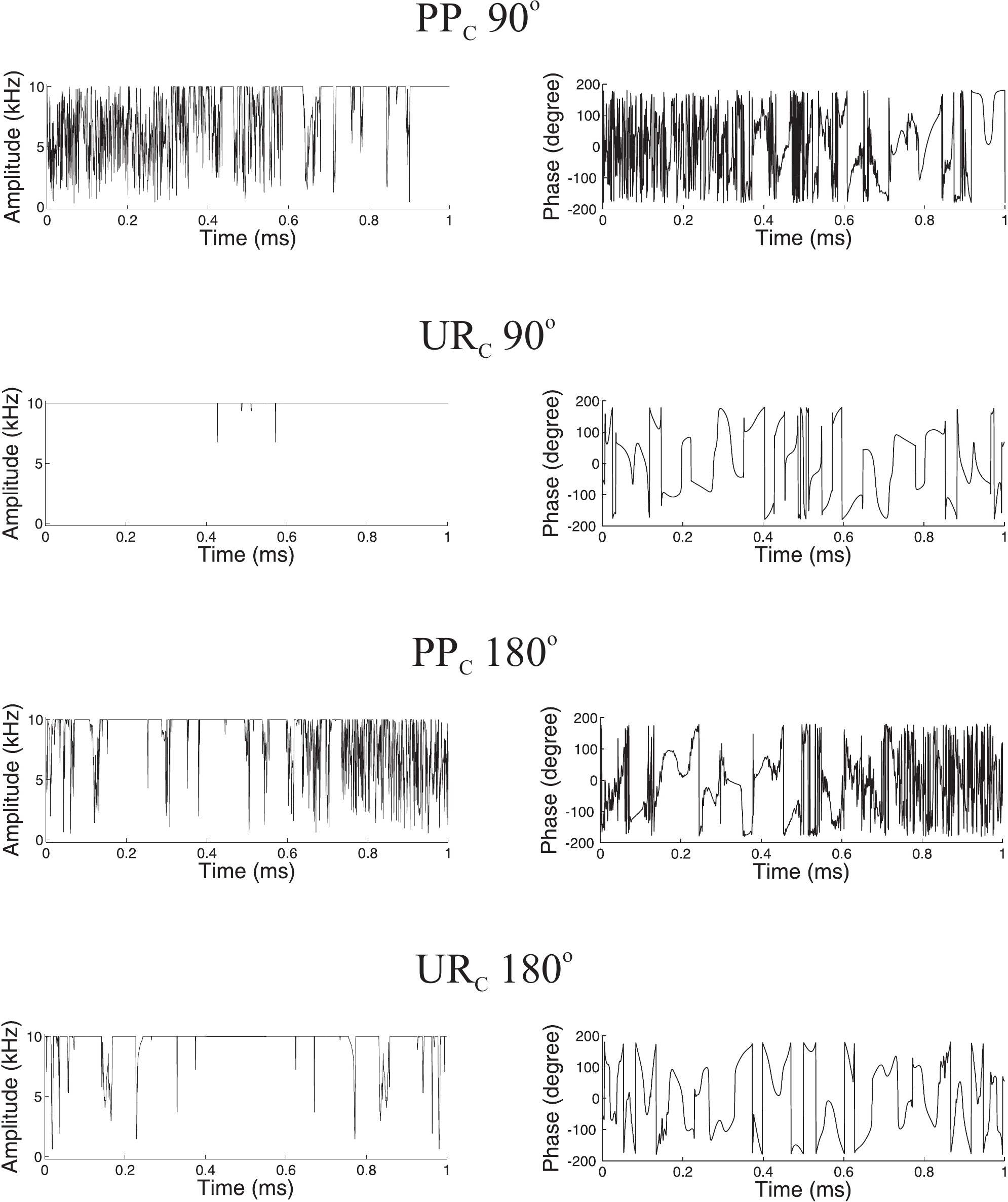}
	\caption[The amplitude and phases of \Cthteen\ Fanta4 pulses.]{The amplitude and phase of \Cthteen\ pulses are plotted as a function of the pulse duration. All are 1~ms long with a maximum amplitude of 10 kHz.}
	\label{fanta_C_pulse}
	\end{figure*}

\begin{figure*}[tb!hp]
\centering
	\includegraphics[width=2\columnwidth]{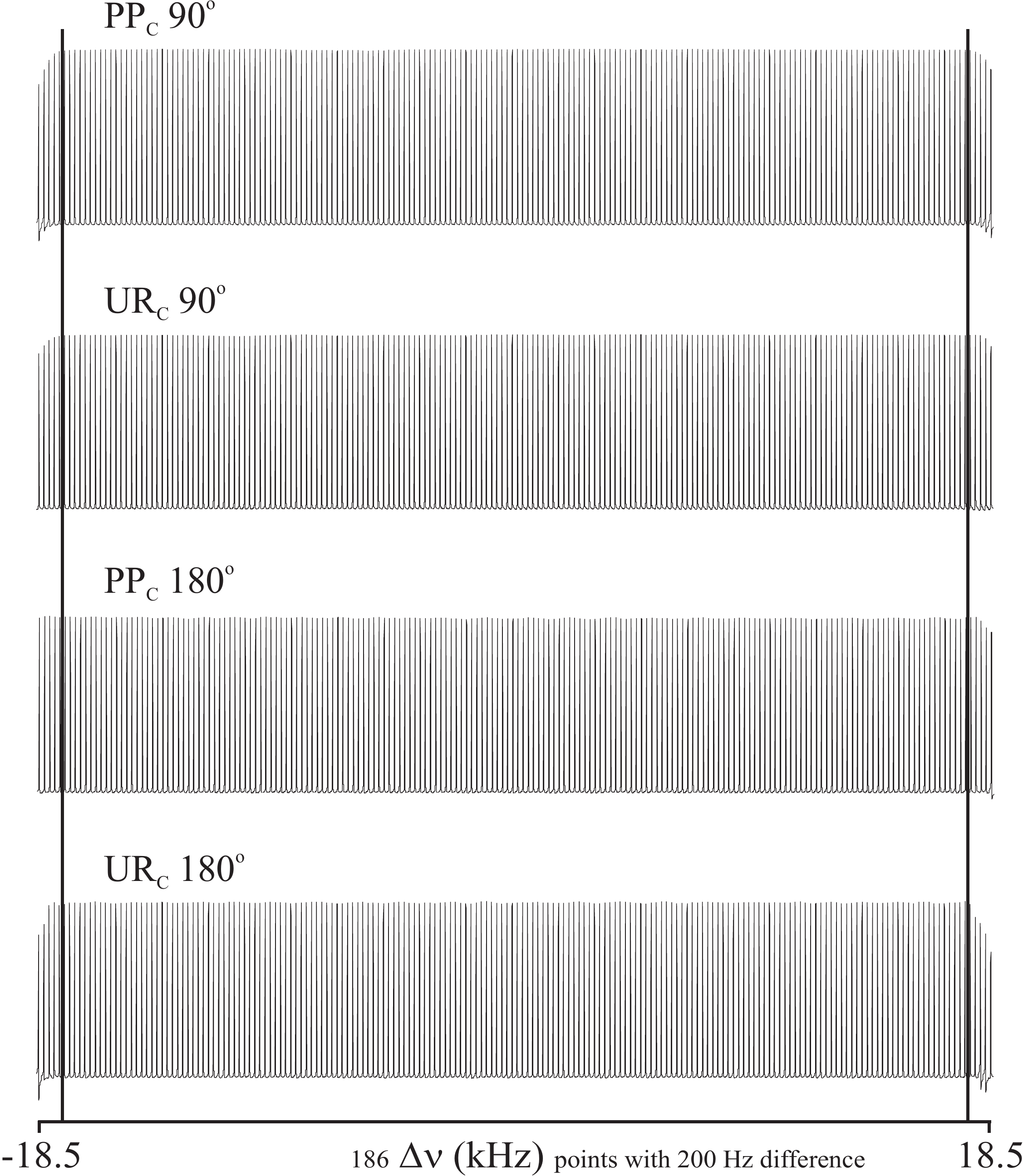}
	\caption [Experimental performance of \Cthteen\ Fanta4 pulses as function of the offset for single spin.] {The experimental performance of \Cthteen\ pulses is plotted as function of 
the offset $\Delta{\nu}$ at ideal RF amplitude with a \RFmax\ of 10 kHz. Experiments were 
aquired on residual proton in HDO for offset range of 37 kHz in steps of 200 Hz. 
\proton\ pulses were optimized for 35 kHz of total offset range with RF miscalibration of 
$\pm5\%$ but the resulting pulses are robust for $\pm10\%$ RF miscalibration. \Cthteen\ PP 90\deg\ and UR 
90\deg\ show the excitation profile for the $z \rightarrow -y$ and \Cthteen\ PP 180\deg\ and UR 180\deg\ show an inversion profile from $z \rightarrow -z$.}
	\label{fig:1d_excitation_profile_fantaCpulse}
	\end{figure*}

All experiments were implemented on a Bruker 600 MHz AVANCE III spectromter equipped with
SGU units for RF control and linearized amplifiers, utilizing a triple-resonance TXI 
probehead and gradients along the z-axis. Measurements are the residual HDO signal using a sample 
of 99.96\% $\rm{D}_2\rm{O}$ doped with ${\rm{CuSO}_4}$ to a $T_1$ relaxation time of 100~ms at 
298$\deg~\rm{K}$. For the \proton\ pulses shown in Figure~\ref{fig:fanta_H_pulse}, signals are obtained for offsets between -11.1 kHz to 11.1 kHz in steps of 200 Hz at an ideal RF amplitude with \RFmax\ of 18 kHz (Fig.~\ref{fig:1d_excitation_profile_fanta_H_pulse}). For the \Cthteen\ pulses shown in Figure~\ref{fanta_C_pulse}, signals are obtained at offsets between -18.5 kHz to 18.5 kHz in steps of 200~Hz at an ideal RF amplitude with \RFmax\ of 10 kHz (Fig.~\ref{fig:1d_excitation_profile_fantaCpulse}). To reduce the effects of RF field inhomogeneity, approximately 40~$\mu$l of sample solution was placed in a 5~mm Shigemi limited volume tube.
The duration of each Fanta4 pulse is 1 ms.

\subsection{Optimization of pulse pairs for two coupled heteronuclear spins 1/2}
\label{twospin}
In the previous section, the  protocol was explained that was necessary to select the set of Fanta4 pulses presented in this paper based on individual pulses optimized for uncoupled spins. Here, we show how the optimal control approach can be used to directly optimize pulse pairs for two $J$-coupled hetero-nuclear spin-1/2 nuclei. However, as this algorithm is computationally demanding, we demonstrate this approach only for a single pulse pair.

\subsubsection{Transfer from single initial to single final state}
\label{twospin_sec1}

The goal is to find a RF pulse which steers the trajectory from a given initial product operator term  {$\rho(0)=2I_\alpha$}{$S_\beta$} at time $t = 0$
to a desired target state {$F=2 I_\gamma$}{$S_\delta$} at $t = T$, where $\alpha, \beta, \gamma,$ and $\delta$ correspond to the $x, y,$ and $z$ components of magnetization.  This is accomplished by optimizing a suitably chosen cost function or performance index, $\Phi$, as discussed further below.

The state of the spin system at time $t$ is characterized by the density operator $\rho(t)$.  The equation of motion in the absence of relaxation is the Liouville-von Neuman equation,
\begin{equation}
\label{leq1}
\dot{\rho}(t) = -i[ ({H_0 +\displaystyle\sum_{k=1}^{m}u_k(t)H_k}), \rho(t)],
\end{equation}
where $H_0$ is the free evolution Hamiltonian, $H_k$ are the RF Hamiltonians corresponding to available control fields, and $u(t) = (u_1(t), u_2(t), ...,u_m(t))$ represents the vector of RF amplitudes, referred to as the control vector.  For the case of two spins considered here, $m = 4$ for the $x$- and $y$-components (or, alternatively, amplitude and phase) of the RF fields applied to the two different spins. The problem is to find the optimal control amplitudes $u_k(t)$ that steer a given initial density operator $\rho(0)$ in a specified time $T$ to a desired target operator $F$ with maximum overlap. 

For Hermitian operators $\rho(0)$ and $F$, this overlap can be measured by the standard inner product
\begin{equation}
\langle F|\rho(T)\rangle = \mathrm{Tr}\{{}F|{\rho(T)}\},
\end{equation}
where the operator Tr returns the trace (sum of diagonal elements) of its argument.  Hence, the performance index, $\Phi$, of the transfer process can be defined as
\begin{equation}
\Phi = \langle F|\rho(T)\rangle.
\label{eq_phi}
\end{equation}
For the full treatment of the optimization procedure we refer to Ref.~\cite{applOCgrape}. 

Digitizing the pulse in $N$ equal steps indexed by $j = 1,2,3,\dots,N$, the basic GRAPE algorithm for this cost is 

\begin{itemize}
\item[1.] Guess initial controls $u_k(j)$.
\item[2.] Starting from $\rho_0$, calculate $\rho(j) = U_j\cdots U_1 \rho_o U_1^{\dagger} \cdots U_j^{\dagger}$ for all $j \le N$.
\item[3.] Starting from ${\lambda_{N}} = {F}$, calculate ${\lambda(j)} =  U_{j+1}^{\dagger} \cdots  U_N^{\dagger} {F} U_N \cdots U_{j+1}$ for all $j \le N$.
\item[4.] Evaluate the gradient $$\delta\Phi/\delta u_k(j)=-\langle\lambda_{j}|i\Delta{t} \lbrack{H_k},\rho_{j}\rbrack \rangle$$ 
and update the $m\times N$ control amplitudes $u_k(j)$.
\item[5.] With these as the new controls, go to step 2.
\end{itemize}

The algorithm is terminated if the change in the performance index $\Phi$ is smaller than a chosen threshold value.

Treatment of non-Hermitian operators and relaxation-optimized coherence transfer can be found in Refs.~\cite{applOCgrape, applOCR}.

\subsubsection{Transfer from two initial to two final states}

Consider a RF pulse which simultaneously transforms two orthogonal initial operators {$\rho_1(0)=2I_\alpha S_\beta$} and  {$\rho_2(0)=2I_\gamma S_\delta$}
to the respective final states {$F_1=2I_\epsilon S_\zeta$} and {$F_2=2I_\eta S_\theta$}, where $\alpha, \beta, \gamma, \delta, \epsilon, \zeta, \eta,$ and $\theta$ correspond as above to the $x, y,$ and $z$ components.  Each state of the spin system is characterized by the density operator $\rho_1(t)$ and $\rho_2(t)$ at time point $t$. The Liouville-von Neuman equation for each state is given by
\begin{equation}
\label{eqLvN}
\dot{\rho}_{n}(t) = -i[ ({H_0 +\displaystyle\sum_{k=1}^{m}u_k(t)H_k}), \rho_{n}(t)],
\end{equation}
where $n = 1, 2, \cdots, P$, labels the states. 

The overall performance index $\Phi$ can be defined as
\begin{equation}
\Phi= {{\langle F_1|\rho_{1}(T)\rangle + \langle F_2|\rho_{2}(T)\rangle}\over {2}}.
\end{equation}

The modified GRAPE algorithm for two states is
\begin{itemize}
\item[1.] Guess initial controls $u_k(j)$

\item[2.] Starting from $\rho_{n0}$, calculate $\rho_{nj} = U_j \cdots U_1~\rho_{n0}~ U_1^{\dagger} \cdots U_j^{\dagger}$  for all $j \le N$  and $n =  1, 2$. 

\item[3.] Starting from ${\lambda_{nN}} = {F_n}$, calculate ${\lambda_{nj}} =  U_{j+1}^{\dagger} \cdots U_N^{\dagger}~{F_n}~U_N \cdots U_{j+1}$ for all $j \le N$ and $n =  1, 2$.

\item[4.] 

Evaluate the gradient $$\delta\Phi/\delta u_k(j)= \hskip 16em$$
$$\hskip 2em -{{\langle\lambda_{1j}|i\Delta{t} \lbrack{H_k},\rho_{1j}\rbrack \rangle+\langle\lambda_{2j}|i\Delta{t} \lbrack{H_k},\rho_{2j}\rbrack \rangle}\over{2}}$$ 
and update the $m\times N$ control amplitudes $u_k(j)$.

\item[5.] With these as the new controls, go to step 2.
\end{itemize}

\begin{figure*}[tb!hp]
\centering
	\includegraphics[width=2\columnwidth]{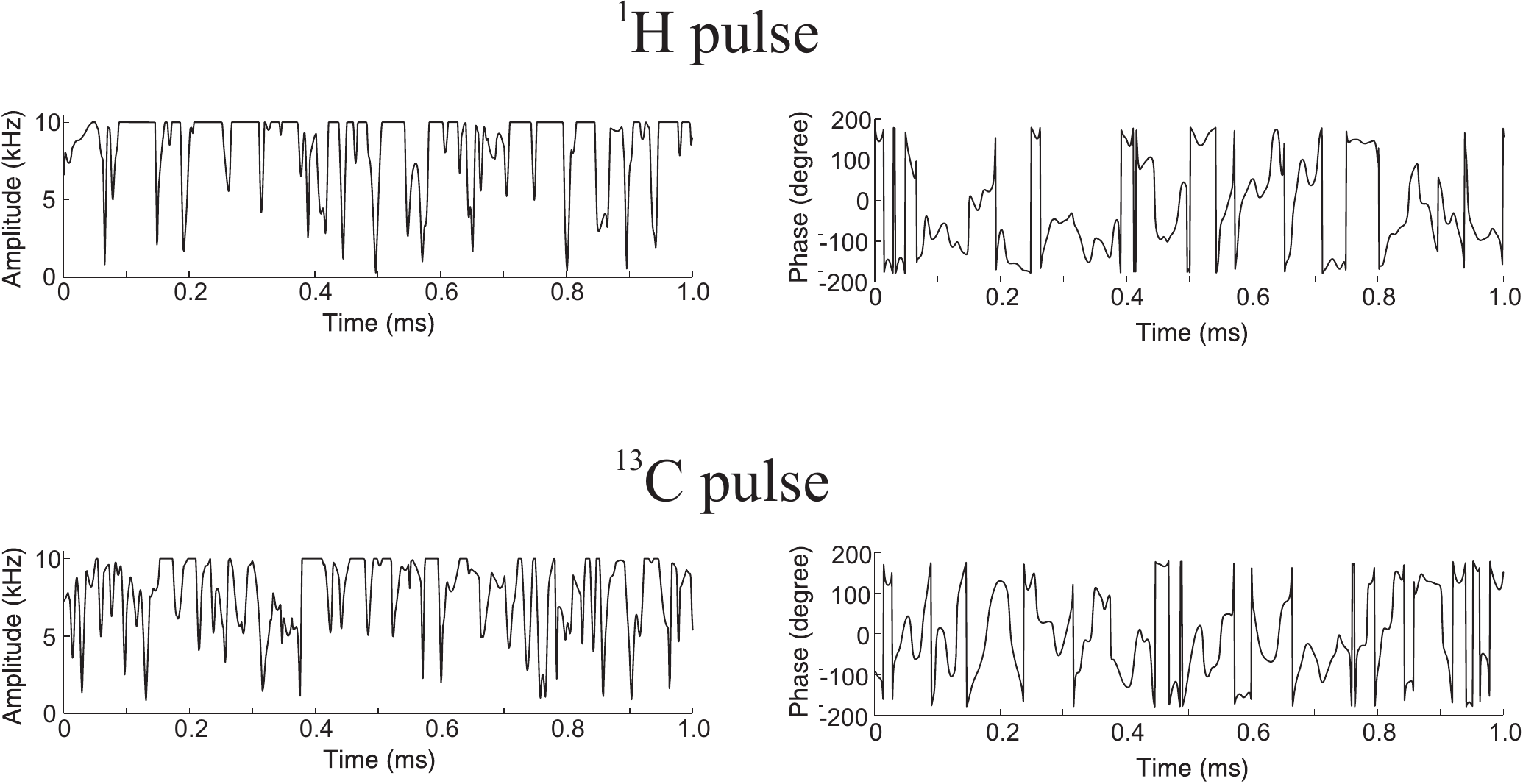}
	\caption[The amplitude and phases of \proton\ and \Cthteen\ pulses optimized simultaneously.]{The amplitude and phase of \proton\ and \Cthteen\ pulses optimized simultaneously for the magnetization transfers
$2 I_z S_x   \rightarrow  2 I_y S_x$ and
$2 I_z S_y \rightarrow  - 2I_y S_z$
	 are plotted as a function of the pulse duration. Both pulses  are 1~ms long with a maximum amplitude of 10 kHz.}
	\label{fig:twoT_pulse}
	\end{figure*}

\begin{figure*}[tb!hp]
\centering
	\includegraphics[width=2\columnwidth]{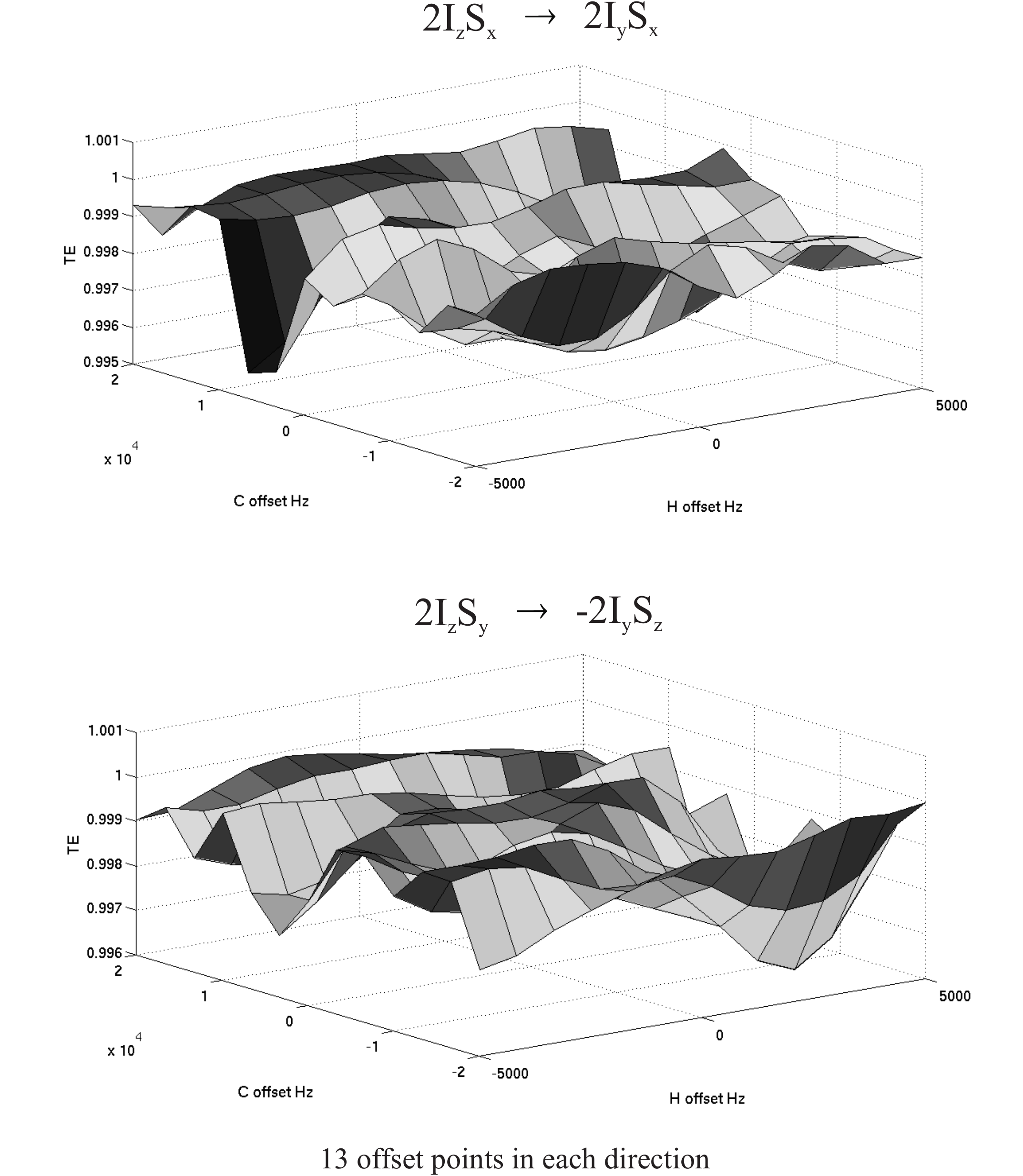}
	\caption [Simulation of \proton\ and \Cthteen\ pulses optimized simultaneously.]{Simulated transfer efficiencies (TE) of the  \proton\ and \Cthteen\ pulses of \Fig{fig:twoT_pulse}, optimized simultaneously for  the magnetization transfers $2 I_z S_x   \rightarrow  2 I_y S_x$ (top) and
$2 I_z S_y \rightarrow  - 2I_y S_z$ (bottom), are plotted as a function of \proton\ and \Cthteen\ offsets.}
	\label{fig:twoT_sim}
	\end{figure*}

Figure~\ref{fig:twoT_pulse} shows the shape of the pulses optimized simultaneously using above algorithm for \proton\ and \Cthteen\ with $J$ coupling of 197~Hz for the following transfers.

\begin{eqnarray}
\label{twpspintransfer}
\rho_1(0)= 2 I_z S_x   &\rightarrow&F1= \ \ 2 I_y S_x \\
\rho_2(0)= 2 I_z S_y &\rightarrow&F2= - 2I_y S_z
\end{eqnarray}
In this example, the maximum pulse amplitudes for both $^1$H and $^{13}$C pulses are 10 kHz and the $^1$H and $^{13}$C offset ranges were $\pm 5$ kHz and $\pm 20$ kHz, respectively.
Figure~\ref{fig:twoT_sim} shows the simulations for the corresponding transfer efficiencies.

\section*{Acknowledgments}
S.J. G. acknowledges support from the DFG (GL 203/6-1), SFB 631 and the Fonds der Chemischen Industrie. M.N. would like to thank the TUM Graduate school T.E.S. acknowledges support from the National Science Foundation under Grant CHE-0943441. The experiments were performed at the Bavarian NMR Center, Technische Universit\"at  M\"unchen.

\bibliographystyle{model1-num-names}
\bibliography{bibtest}

	

	


\end{document}